\documentclass[twocolumn,aps,prb,showpacs,amsmath,amssymb,10pt]{revtex4-1}
\usepackage{amsmath}
\usepackage{cases}
\usepackage{amssymb}
\usepackage{amsfonts}
\usepackage{txfonts}
\usepackage[dvips]{graphicx}
\usepackage{bm}
\usepackage{color}
\usepackage{bbold}

\begin{document}

\title{Anderson localization in two-dimensional graphene with short-range disorder: 
One-parameter scaling and finite-size effects}
\author{Zheyong Fan}
\email[Corresponding author: ]{zheyong.fan@aalto.fi}
\author{Andreas Uppstu}
\author{Ari Harju}
\affiliation{COMP Centre of Excellence,
Department of Applied Physics, Aalto University, Helsinki, Finland}
\date{\today}

\begin{abstract}
We study Anderson localization in graphene with short-range disorder using the real-space Kubo-Greenwood 
method implemented on graphics processing units. Two models of short-range disorder, namely, 
the Anderson on-site disorder model and the vacancy defect model, are considered. For graphene with Anderson disorder, 
localization lengths of quasi-one-dimensional systems with various disorder strengths, edge symmetries, 
and boundary conditions are calculated using the real-space Kubo-Greenwood formalism, showing excellent 
agreement with independent transfer matrix calculations and superior computational efficiency. 
Using these data, we demonstrate the applicability of the one-parameter scaling theory of localization 
length and propose an analytical expression for the scaling function, which provides a reliable method 
of computing the two-dimensional localization length. This method is found to be consistent with another 
widely used method which relates the two-dimensional localization length to the elastic mean free path and the 
semiclassical conductivity. Abnormal behavior at the charge neutrality point is identified and 
interpreted to be caused by finite-size effects when the system width is comparable to or smaller than the 
elastic mean free path. We also demonstrate the finite-size effect when calculating the two-dimensional 
conductivity in the localized regime and show that a renormalization group beta function consistent 
with the one-parameter scaling theory can be extracted numerically. For graphene with vacancy disorder, 
we show that the proposed scaling function of localization length also applies.
Lastly, we discuss some ambiguities in calculating the semiclassical conductivity around the 
charge neutrality point due to the presence of resonant states.
\end{abstract}

\maketitle

\section{Introduction}

Graphene is an effectively two-dimensional (2D) material consisting of a sheet of carbon atoms. 
\cite{geim2007, katsnelson2012} In its pristine form, it exhibits many remarkable low-energy 
electronic transport properties, such as half-integer quantum Hall effect \cite{novoselov2005,zhang2005} 
and Klein tunneling, \cite{katsnelson2006} due to the linear dispersion of the charge carriers 
near two inequivalent valleys around the charge neutrality point. However, disorder may 
dramatically alter both the electronic structure \cite{neto2009} and transport properties 
\cite{peres2010,mucciolo2010,sarma2011} of graphene. It is generally believed that both short-range 
\cite{aleiner2006,altland2006,xiong2007,schubert2010} and strong long-range \cite{zhang2009} disorder can lead 
to inter-valley scattering and Anderson localization, while weak long-range disorder only gives 
rise to intra-valley scattering, which does not lead to backscattering and Anderson localization 
\cite{ostrovsky2007,bardarson2007,nomura2007}.

Due to its intrinsic low-dimensionality, graphene provides an ideal testbed of revisiting old 
ideas regarding Anderson localization in low dimensions as well as discovering new ones. The 
most successful theory for Anderson localization is one-parameter scaling \cite{abrahams1979}, 
which predicts that all states in disordered one- and two-dimensional systems are localized at 
zero temperature if the system is sufficiently large, although exceptions can occur when the 
disorder is correlated \cite{rodriguez2012} or electron-electron interaction cannot be 
neglected \cite{punnoose2005}. However, recent works regarding localization in graphene have 
yielded results that conflict with one-parameter scaling, with some studies supporting the 
existence of mobility edges even in the presence of uncorrelated Anderson disorder 
\cite{amini2009, song2011}. Very recent numerical results indicate the difficulty of associating 
data for the finite-size localization length with a single scaling curve \cite{gonzalez2013} 
as well as the discrepancy between results of the 2D localization length obtained from the 
finite-size scaling approach and the self-consistent theory of localization \cite{lee2013}. 
On the other hand, it has been suggested that the conductivity at the charge neutrality point (CNP) saturates to 
a constant value \cite{ostrovsky2010}, or decays following a power-law rather than exponentially 
with increasing system size \cite{cresti2013,laissardiere2013},  
in graphene with resonant scatterers such as vacancy defects. 

Since the typical length scales regarding localization properties in 2D systems are generally 
very large, efficient numerical methods are desirable. Although the standard numerical method 
for studying quantum transport is the Landauer-B\"{u}ttiker approach combined with the recursive 
Green-function technique, using it for realistically sized truly 2D graphene systems is still 
beyond current computational ability, since the computational effort scales cubically with the 
width of the system. In contrast, the linear-scaling real-space Kubo-Greenwood (RSKG) method 
\cite{mayou1988,mayou1995,roche1997,triozon2002} is generally much more efficient and has been 
used to study electronic transport in realistically sized graphene with various kinds of disorder 
\cite{cresti2013,laissardiere2013,lherbier2008,leconte2011,lherbier2012,radchenko2012,botello2013}. 
In this method, the actual computational effort depends on the energy resolution, the required 
statistical accuracy, and most crucially, the transport regime. Exploring the localization 
properties generally requires a large simulation cell to eliminate possible finite-size effects 
and a long correlation time (which can be thought as the evolution time of a wavepacket) to 
actually reach the localized regime, which can be very time-consuming. Recently, we have 
significantly accelerated the calculations by implementing \cite{fan2014} this method on graphics 
processing units \cite{harju2013}, and further developed methods for obtaining the localization 
properties of disordered systems. It has been established \cite{uppstu2014} through comparisons 
with the standard Landauer-B\"{u}ttiker approach that (1) the average propagating length of 
electrons can serve as a good definition of length before its saturation and (2) the saturated 
propagating length is directly proportional to the localization length defined in terms of the 
exponential decay of conductance in the strongly localized regime.

Armed with this efficient numerical method, we perform an extensive numerical 
study of Anderson localization in graphene with short-range disorder, including Anderson 
disorder and vacancies. We first calculate the localization lengths for various 
quasi-one-dimensional (Q1D) systems using the RSKG method. 
Since most of the previous works \cite{xiong2007,gonzalez2013,lee2013,schreiber1992} have 
applied the transfer matrix method (TMM) \cite{pichard1981} (or equivalently, the recursive 
Green-function method, see Ref. \onlinecite{mackinnon1983}), we also present a comparison 
between these two methods. Based on our computational data, we are able to compare the 
results against the one-parameter scaling theory of localization length 
\cite{mackinnon1983,mackinnon1981} and construct  an analytical expression 
for the so far undetermined scaling function. Our results are consistent with those
of Schreiber and Ottomeier \cite{schreiber1992} and Lee \textit{et al.} \cite{lee2013}, 
but compared to these works, we have considered a more complete set of energy points and much wider 
systems. We will also discuss the finite-size effects for the scaling analyses 
of both localization length and conductivity and some ambiguities in determining the semiclassical 
conductivity in graphene with resonant disorder using the RSKG method.

This paper is organized as follows. Section \ref{section:methods} defines the physical models
and introduces the TMM for the calculation of localization length 
and the RSKG method for the calculation of localization length as well as
other electronic and transport properties. We then study Anderson localization of
graphene with Anderson disorder and vacancy-type disorder in Sections \ref{section:anderson}
and \ref{section:vacancy}, respectively. Section \ref{section:conclusions} concludes.

\section{\label{section:methods} models and methods}

\subsection{Models}

For pristine graphene, we apply the widely used nearest-neighbor $p_z$ orbital tight-binding Hamiltonian
\begin{equation}
H = -t\sum_{\langle i, j\rangle} |i\rangle \langle j|
\end{equation}
where $t$ is the hopping parameter. The uncorrelated Anderson disorder is modeled by adding 
random on-site potentials uniformly distributed within an energy interval of $[-W/2, W/2]$, 
$W$ being a measure of the disorder strength. The more realistic vacancy disorder is modeled 
by randomly removing carbon atoms according to a prescribed defect concentration $n$, which 
is defined to be the number of vacancies divided by the number of carbon atoms in the pristine 
system. We will consider the whole energy spectrum for the Anderson model and thus take $t$ as 
the unit of energy, but only consider a small energy window for the vacancy model and take eV 
as the unit of energy and set $t=2.7$ eV. When calculating the Q1D localization length, we will 
consider both zigzag and armchair graphene nanoribbons (ZGNRs and AGNRs, correspondingly).
To test the effect of the boundary conditions in the transverse direction, we also consider
armchair carbon nanotubes (ACNTs) with the transport direction along the zigzag edge and 
periodic boundary conditions also along the transverse direction. We use $N_x$ and $N_y$ to denote 
the number of dimer lines along the zigzag edge and the number of zigzag-shaped chains across 
the armchair edge, respectively. The total number of carbon atoms in the computational cell
is then $N_x \times N_y$. The symbol $M$ defines the width of the system. 
For ZGNRs and ACNTs, we set $M$ to $N_y$ and obtain the actual width $L_M$ using $L_M=3Ma/2$. 
For AGNRs, we set $M$  to $N_x$ and obtain the actual width using $L_M=\sqrt{3}Ma/2$. 
Here, $a$ is the carbon-carbon distance, being roughly 0.142 nm.

\subsection{Methods}

We define the localization length $\lambda_M$ of a Q1D system with a fixed width $L_M$ 
to be the characteristic length of the exponential decay of typical conductance with 
the system length $L$ in the strongly localized regime \cite{note_on__convention}:
\begin{equation}
\label{equation:xi_from_g}
g_{\textmd{typ}}(L) \sim \exp(-2L/\lambda_M),
\end{equation}
where the typical conductance $g_{\textmd{typ}}\equiv \exp(\langle \ln g \rangle)$
is obtained from the ensemble average 
over individual realizations with 
fixed system size and disorder strength \cite{anderson1980}.

In the literature, the most often used methods for computing $\lambda_M$ are the 
recursive Green-function method and the TMM, which are essentially equivalent \cite{mackinnon1983}. 
In Ref. \onlinecite{uppstu2014}, we have suggested another method of finding $\lambda_M$ 
using the RSKG formalism, briefly explained below. In this work, we will further 
demonstrate its accuracy and efficiency by comparing it against the TMM.

\subsubsection{The transfer matrix method}

In the TMM, the wave function $\psi_n$ of the $n$th 
slice along the transport direction of 
the Q1D geometry is calculated iteratively using the transfer matrix equation 
(note that all the matrix or vector elements here are $M$-by-$M$ matrices) as
\begin{equation}
\left( \begin{array}{cc}
\psi_{n+1} \\
\psi_{n} \end{array} \right)=
\left( \begin{array}{cc}
E \mathbb{1}-H_{n} & -\mathbb{1} \\
\mathbb{1} & \mathbb{0}  \end{array} \right)
\left( \begin{array}{cc}
\psi_{n} \\
\psi_{n-1} \end{array} \right)
\equiv T_n
\left( \begin{array}{cc}
\psi_{n} \\
\psi_{n-1} \end{array} \right),
\end{equation}
with the initial wave functions $\psi_1 = \mathbb{1}$ and $\psi_0=\mathbb{0}$. 
We only consider ZGNRs and ACNTs (both with the transport direction along the zigzag edge) 
when using the TMM, where the matrix $H_n$ takes two alternative forms depending on whether 
$n$ is even or odd, as given in Ref. \onlinecite{schreiber1992}. According to Oseledec's 
theorem \cite{oseledec1968}, with increasing $N$, the eigenvalues of 
$\left(\Gamma_N^{\dagger}\Gamma_N\right)^{1/2N}$, where $\Gamma_N \equiv T_NT_{N-1} \cdots T_1$, 
converge to fixed values $e^{\pm \gamma_m}$, the $\gamma_m~(1\leq m \leq M)$ being Lyapunov exponents. 
The localization length is defined as the largest decaying length associated with the minimum 
Lyapunov exponent \cite{note_on__convention}:
\begin{equation}
\label{equation:xi_from_gamma}
\lambda_M = \frac{1}{\gamma_{\textmd{min}}}.
\end{equation}
Numerically, the minimum Lyapunov exponent can be computed by combining Gram-Schmidt 
orthonormalization with the above transfer matrix multiplication. Practically, 
only sparse matrix-vector multiplication is required and one does not need to perform 
Gram-Schmidt orthonormalization after each multiplication. Usually, performing one Gram-Schmidt 
orthonormalization every ten multiplications keeps a good balance between speed and accuracy. 
The number of slices required for achieving a relative accuracy of $\epsilon$ is approximately 
\cite{mackinnon1983} $2(\lambda_M/a)/\epsilon^2$. In this work, we set $\epsilon=1\%$.

\subsubsection{The real-space Kubo-Greenwood method}
In the RSKG method \cite{mayou1988, mayou1995, roche1997, triozon2002}, the zero-temperature dc 
electrical conductivity at energy $E$ and correlation time $\tau$ can be expressed as
\begin{equation}
\label{equation:rec}
\sigma(E, \tau) = e^2 \rho(E) \frac{d\Delta X^2(E, \tau)}{2d\tau},
\end{equation}
where
\begin{equation}
\rho(E) = \frac{2\textmd{Tr}\left[\delta(E-H)\right]}{\Omega}
\end{equation}
is the electronic density of states with the spin degeneracy taken into account. 
Note that the factors of 2 in the above two equations can cancel each other and are not presented 
in some works, but we prefer to keep them for clarity. Here, $H$ is the Hamiltonian and $\Omega$ is the volume, 
or in our case, just the area of the graphene sheet, and
\begin{equation}
\label{equation:msd}
\Delta X^2(E, \tau) = \frac{\textmd{Tr}\left[[X, U(\tau)]^{\dagger} \delta(E-H) [X, U(\tau)]\right]}
{ \textmd{Tr}\left[\delta(E-H)\right]}
\end{equation}
is the mean square displacement. $X$ is the position operator and $U(\tau)=e^{-iH\tau/\hbar}$ 
is the time-evolution operator. What need to be calculated are $\textmd{Tr}\left[\delta(E-H)\right]$ and
$\textmd{Tr}\left[[X, U(\tau)]^{\dagger} \delta(E-H) [X, U(\tau)]\right]$ at a chosen set of $\tau$. 
The so-called linear-scaling algorithm for calculating the latter (the calculation of the former 
does not need the second technique below) can be achieved by the following three techniques: 
(1) approximating the trace by using one or a few random vectors $|\phi\rangle$, 
$\textmd{Tr}[A]\approx\langle \phi|A|\phi\rangle$, $A$ being an arbitrary operator, 
(2) calculating the time-evolution of $[X, U(\tau)]|\phi\rangle$ iteratively using, e.g., 
the Chebyshev polynomial expansion, and (3) approximating the Dirac delta function $\delta(E-H)$ 
using a linear-scaling technique such as Fourier transform, Lanczos recursion, 
or kernel polynomial. The relative error caused by the random-vector approximation
is proportional to  \cite{weisse2006} $1/\sqrt{N_r N}$, where $N$ is the Hamiltonian size 
(the total number of carbon atoms in our problems) and $N_r$ is the number of 
independent random vectors used. In this work, we have used a few to a few tens of random
vectors for each simulated system, the specific number depending on the specific system,
the required accuracy, and the specific quantities 
to be calculated. For the approximation of the Dirac delta function, we have used
the kernel polynomial method \cite{weisse2006}. The energy resolution $\delta E$ achieved using
this method is inversely proportional to the number of Chebyshev moments (which is the order 
the Chebyshev polynomial expansion) $N_m$ used. For most of the calculations, 
we have chosen $N_m$ to be 3000, which corresponds to en energy resolution of a few meV.
While this energy resolution is sufficiently high for graphene
with Anderson disorder, it is not neccesarily high enough to distinguish the resonant
state at the CNP in graphene with vacancy defects from other states. 
In Section \ref{section:resolution}, we will disscuss
the effect of energy resolution on the results for graphene with vacancy defects.
Details of the involved algorithms and the implementation on graphics processing units 
can be found in Ref. [\onlinecite{fan2014}].

As $\tau$ increases from zero, the running conductivity $\sigma(E, \tau)$ first increases linearly, 
indicting a ballistic behavior, and then gradually saturates to a fixed value, which can be 
interpreted as the semiclassical conductivity $\sigma_{\textmd{sc}}(E)$, and finally decreases 
until it becomes zero if localization takes place. In practice, especially when the disorder is strong, 
there may be no apparent plateau to which the running conductivity saturates, 
and $\sigma_{\textmd{sc}}(E)$ is thus usually defined as the maximum of $\sigma(E, \tau)$. 
While this is generally a reasonable definition, it can sometimes result in problems, 
as we will show in Section \ref{section:vacancy_b}. 
After obtaining $\sigma_{\textmd{sc}}(E)$, one can calculate the elastic mean free path $l_\textmd{e}(E)$ 
through the Einstein relation for diffusive transport \cite{beenakker1991}:
\begin{equation}
\label{equation:mfp}
\sigma_{\textmd{sc}}(E) = \frac{1}{2}e^2 \rho(E) v(E) l_\textmd{e}(E),
\end{equation}
where $v(E)$ is the Fermi velocity, which can be calculated from the velocity 
auto-correlation at zero correlation time \cite{fan2014}.

The usefulness of the RSKG method also depends crucially on a definition of propagating 
length $L(E, \tau)$ in terms of $\sqrt{\Delta X^2(E, \tau)}$. Indeed, in the original 
Kubo-Greenwood formalism, there is no definition of length and no connection between 
conductivity and conductance can be made. A definition of length is required for the study 
of mesoscopic transport properties. A natural definition would be $L(E, \tau) = \sqrt{\Delta X^2(E, \tau)}$, 
but a more precise relation has been established \cite{fan2014,uppstu2014}:
\begin{equation}
\label{equation:length}
 L(E, \tau) = 2 \sqrt{\Delta X^2(E, \tau)}.
\end{equation}
The factor of 2 in this equation can be justified from different perspectives: (1) it results 
in \cite{fan2014} the textbook formula \cite{datta2012} for the ballistic conductance
\begin{equation}
\label{equation:ballistic_conductance}
g(E)=e^2\rho(E)v(E)L_M/2,
\end{equation}
and (2) it results in a Q1D conductance $g(E, L)=L_M \sigma(E, \tau)/L(E,\tau)$ which is 
consistent with independent Landauer-B\"uttiker calculations in the localized regime \cite{uppstu2014}. 
This definition of length 
is only valid up to about $g\sim 0.1e^2/h$, 
after which the propagating length saturates  
to a fixed value proportional to the localization length
\cite{uppstu2014,note_on__convention}:
\begin{equation}
\label{equation:xi_from_msd}
 \lambda_M(E) = \lim_{\tau\rightarrow \infty} \frac{2\sqrt{\Delta X^2(E,\tau)}}{\pi}.
\end{equation}
The meaning of the factor of $\pi$ in this equation is yet to be found, but this expression yields 
results in a good agreement with independent Laudauer-B\"uttiker calculations \cite{uppstu2014}. 
Although an infinite $\tau$ is indicated in the above equation, in practice, we only simulate up 
to a finite $\tau$ and then fit the mean square displacement data using a Pad\'e approximant of the form
$\Delta X^2(\tau)=(c_1\tau+c_2)/(\tau+c_3)$. We found that as long 
as the mean square displacement is almost converged, 
this simple Pad\'e approximant results in a very good fit to the data and the saturated mean square 
displacement can be extracted as $c_1$. 
As in the case of the TMM, an error estimation of the calculated data is useful to 
evaluate the quality of the results. However, there seems to be no 
unique way to define the errors for $\lambda_M(E)$ calculated using the RSKG method. 
We have estimated the error for $\lambda_M(E)$ as the mean of
$|L(E, \tau)-L_{\textmd{fit}}(E, \tau)|$ over $\tau$, where $L_{\textmd{fit}}(E, \tau)$
is the fitted propagating length using the Pad\'e approximant. 
We will further validate this method by comparing with 
independent TMM calculations in Section \ref{section:anderson_a} and discuss the 
finite-size effect in this method caused by the finite simulation cell length 
in Section \ref{section:vacancy_a}.

\section{\label{section:anderson}Graphene with Anderson disorder}

\subsection{\label{section:anderson_a}Localization lengths for quasi-one-dimensional systems}

\begin{figure*}[htb]
\begin{center}
  \includegraphics[width=8cm]{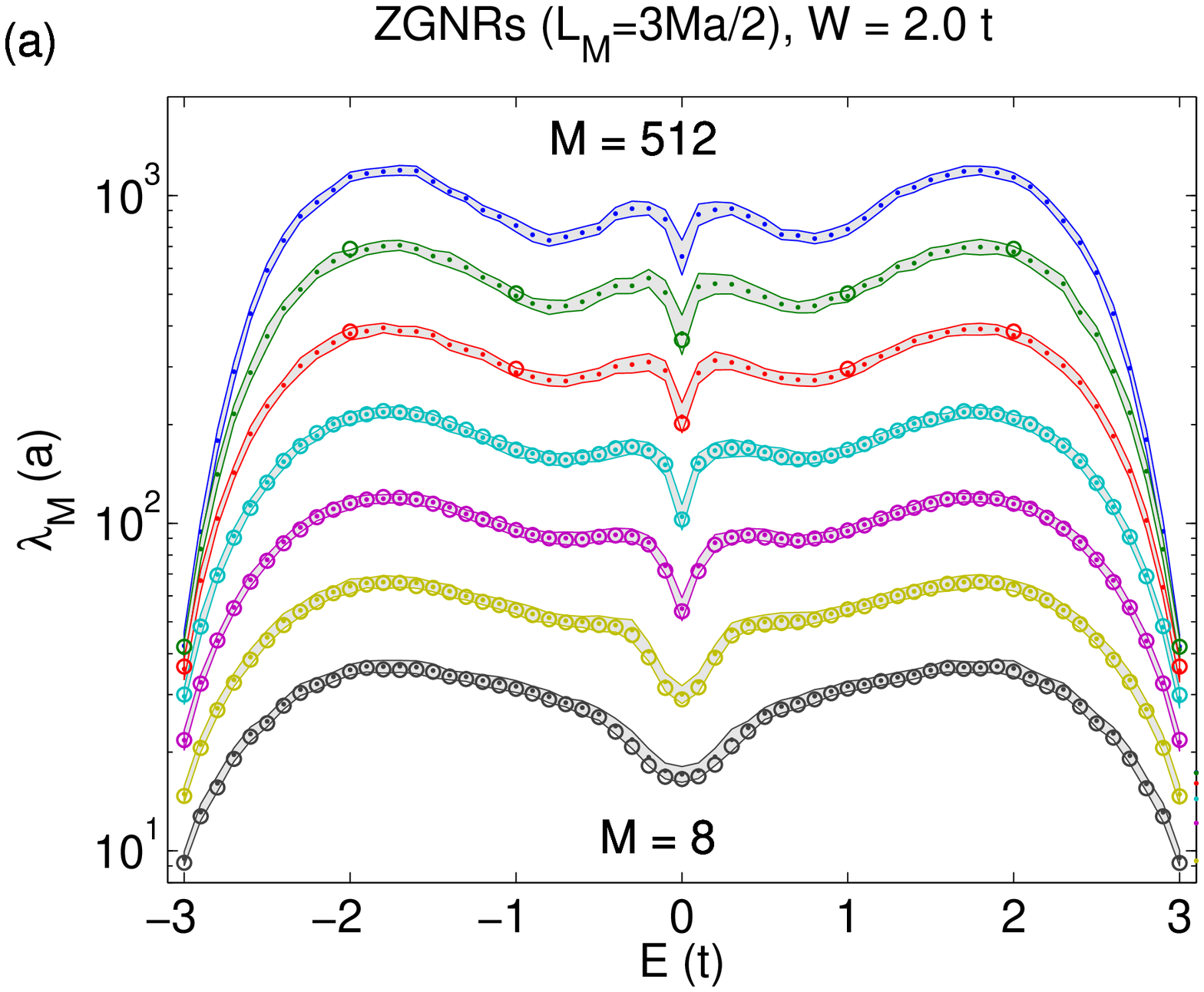}
  \includegraphics[width=8cm]{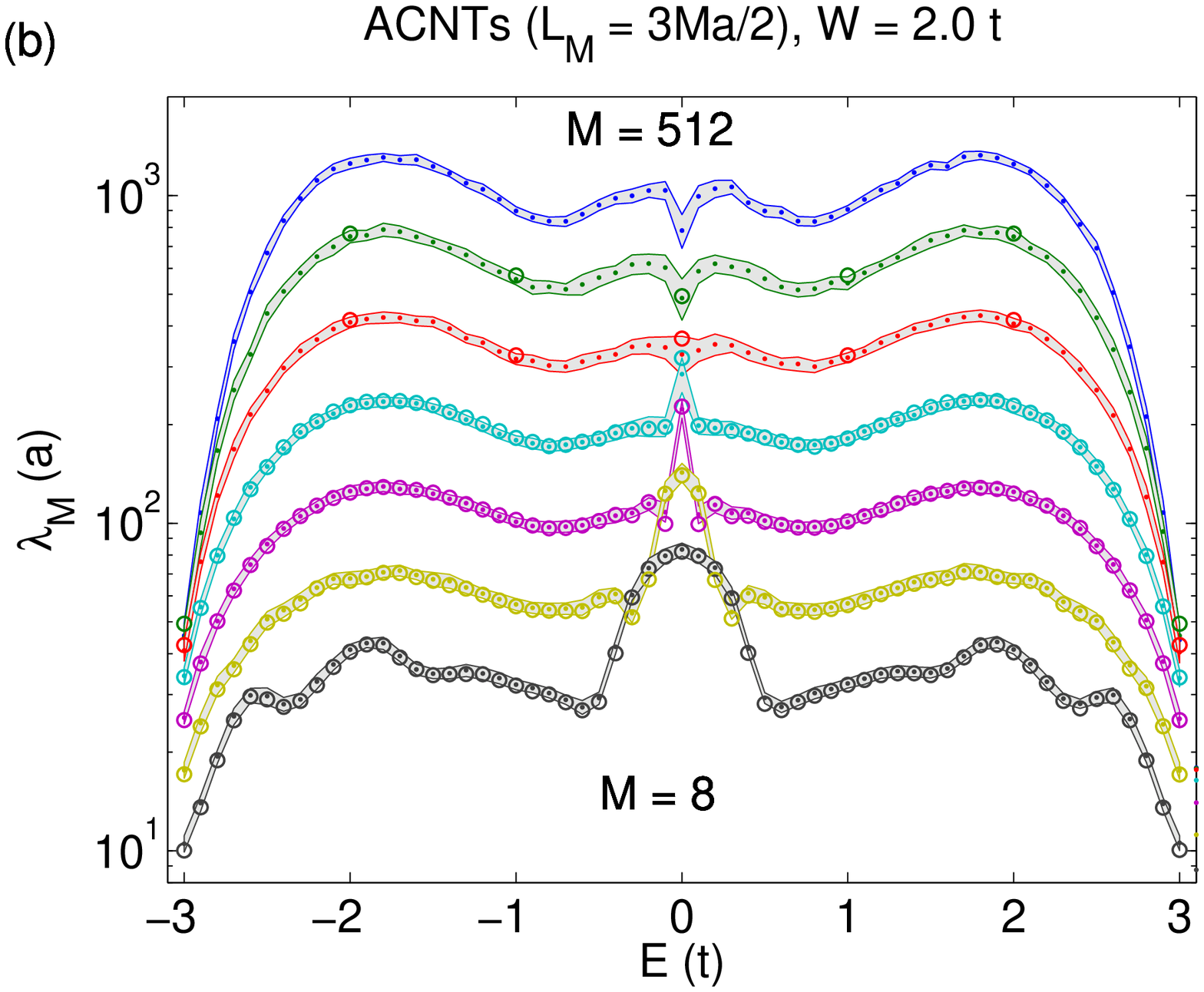}\\
  \includegraphics[width=8cm]{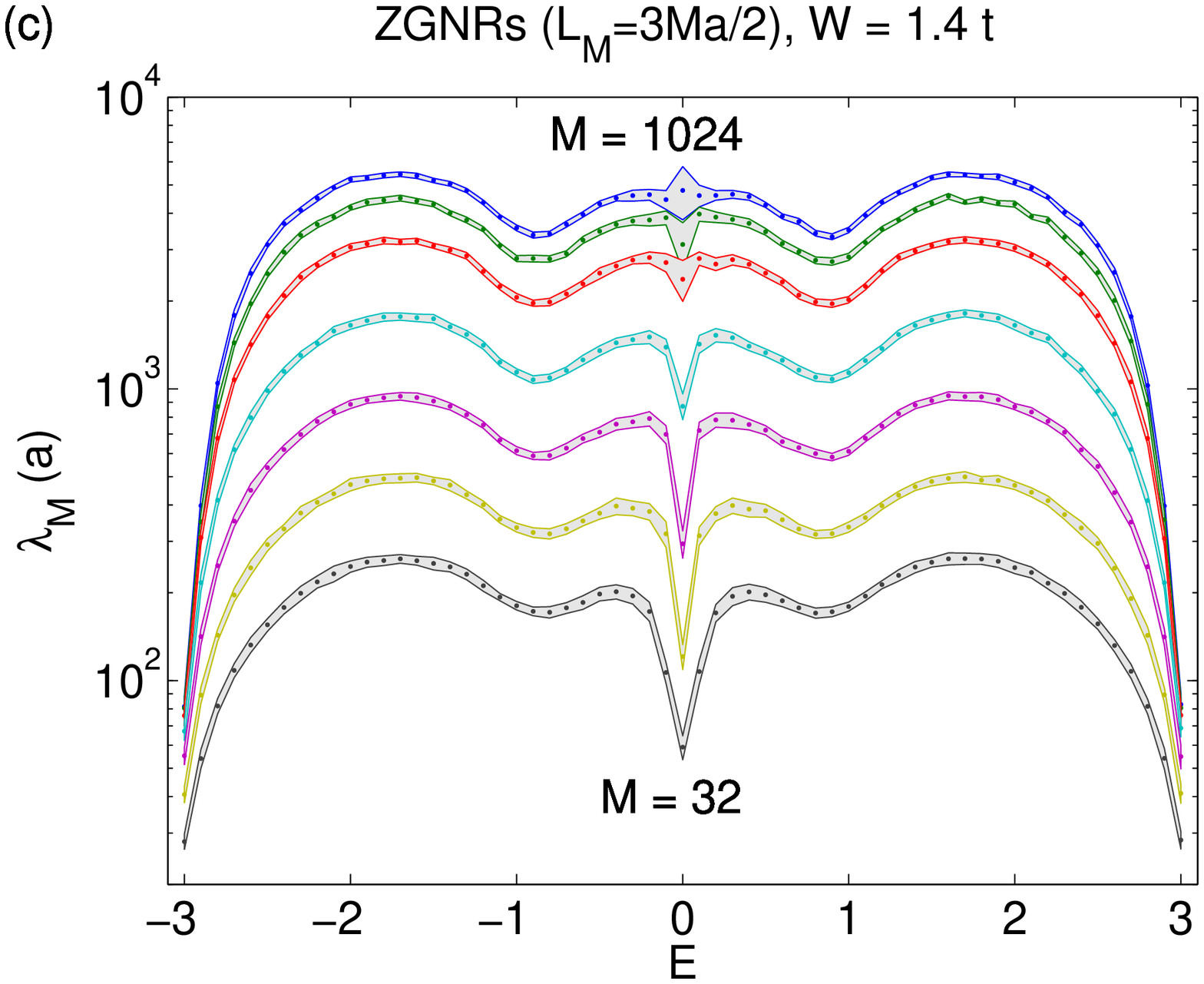}
  \includegraphics[width=8cm]{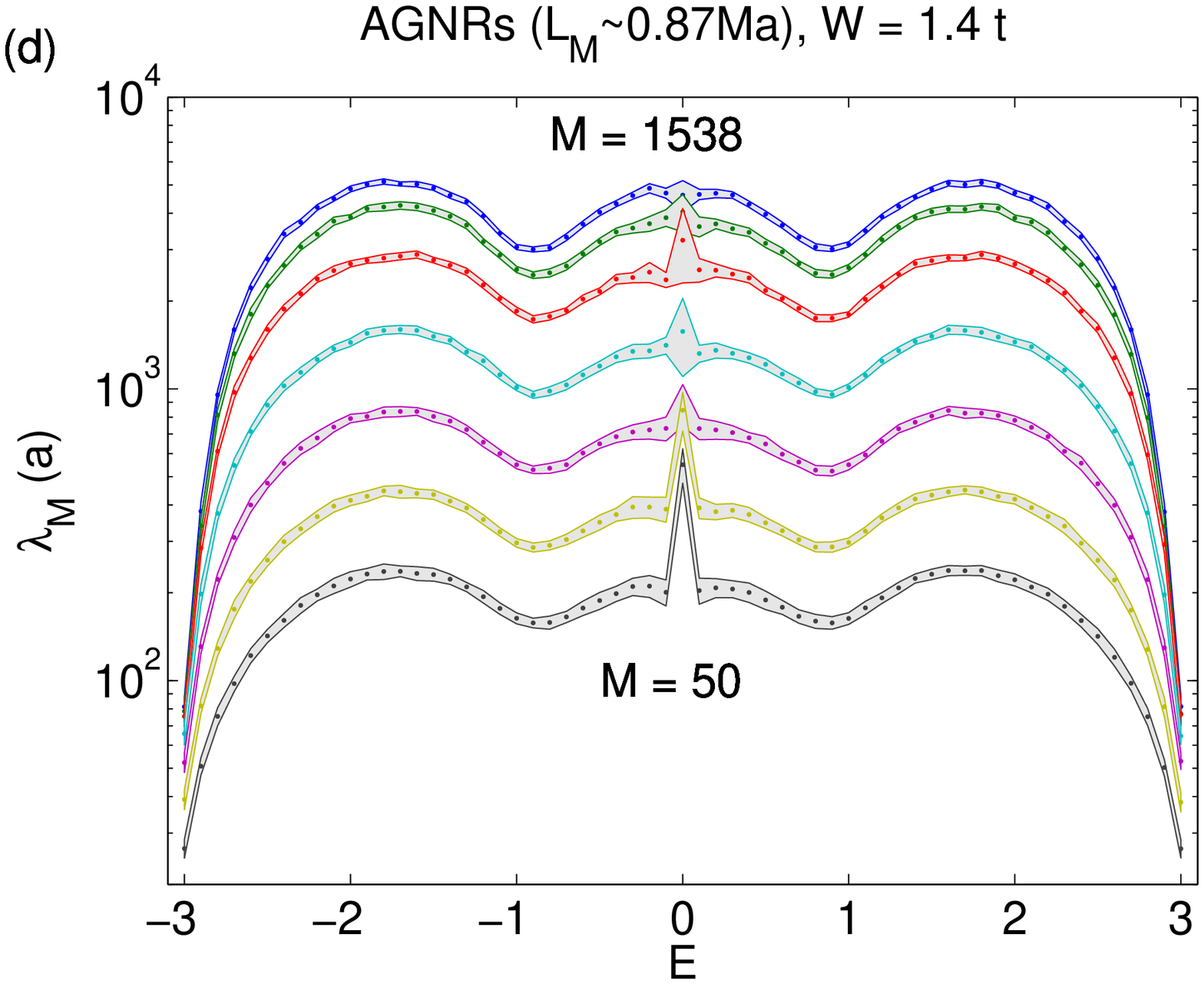}\\
  \caption{(color online)
Localization lengths as a function of energy for Q1D systems: 
(a) ZGNRs with $W=2t$, (b) ACNTs with $W=2t$, (c) ZGNRs with $W=1.4t$, and (d) AGNRs with $W=1.4t$. 
For (a) and (b), $M=$ 8, 16, 32, 64, 128, 256, and 512; for (c), $M=$ 32, 64, 128, 256, 512, 768, and 1024; 
for (d), $M=$ 50, 98, 194, 386, 770, 1154, and 1538. The open circles (only in (a) and (b)) and the small solid dots
represent the results obtained by the TMM and the RSKG method, respectively. The shaded areas with bounding lines
indicate the error estimates of the data calculated by the RSKG method.
The value of $M$ increases monotonically from bottom to top in each subfigure.
Note the different relation between the width $L_M$ and $M$ for AGNRs from other cases. }
\label{figure:lambda_M}
\end{center}
\end{figure*}

Figure 1 shows the calculated localization lengths for Q1D systems with different widths, 
energies, disorder strengths, edge types, and boundary conditions. The considered systems are 
(a) ZGNRs with $W=2.0t$, (b) ACNTs with $W=2.0t$, (c) ZGNRs with $W=1.4t$, and (d) AGNRs with $W=1.4t$. 
In Figs. 1(a) and 1(b), the open circles and small filled dots correspond to the results obtained by the 
TMM and the RSKG method, respectively. The errors estimates for the RSKG results
are indicated by the shaded areas with bounding lines. 
The relative accuracy of the TMM results is set to $1\%$, which
would result in errors comparable to the corresponding marker size, and we thus
omit the error bars for the TMM results for simplicity.
Both methods give practically the same results, but the 
RSKG method is much more efficient for wider systems due to the use of linear scaling techniques 
and the intrinsic parallelism in energy of this method. The parallelism in energy means that 
obtaining the results for all the energy points does not require more computation time than 
obtaining the result for a single energy value. In contrast, the computation time for the TMM 
scales cubically with respect to the width of the system and there is no parallelism in energy. 
Therefore, using the TMM, we have only calculated a limited number of energy 
points for $M=128$ and 256 and no points for $M=512$. Even under these conditions, 
the computation times for these two methods are roughly equal, which demonstrates 
the accuracy and efficiency of the RSKG method. We thus only used the RSKG method for weaker 
disorder, as shown in Figs. 1(c) and 1(d).

There is an obvious difference between the results for different boundary conditions and edge types. 
Figures 1(a) and 1(b) correspond to transport in the direction of the zigzag edge, and differ 
only by the boundary conditions used in the transverse direction, with Fig. 1(a) corresponding to 
free (hard wall) boundary conditions (ZGNRs) and Fig. 1(b) to periodic boundary conditions (ACNTs). 
We note that for ACNTs, the CNP behaves rather differently from the other points: 
it evolves from a local maximum for $M<128$ to a local minimum for $M>128$. This observation is 
consistent with the finding by Xiong \textit{et al.} \cite{xiong2007}. 
Figures 1(c) and 1(d) correspond to a weaker disorder with $W=1.4t$, 
with Fig. 1(c) showing results for ZGNRs and 1(d) for AGNRs.
To avoid band gaps, only metallic AGNRs are considered. We note that 
AGNRs behave similarly as ACNTs, having a maximum of $\lambda_M$ at the CNP when the width of the system 
is small. However, with increasing width, the differences between different boundary conditions 
and edge types become smaller, and one may expect that these differences become vanishingly small 
in the limit of wide systems.

\subsection{One-parameter scaling of localization length}

\begin{figure*}[htb]
\begin{center}
  \includegraphics[width=16cm]{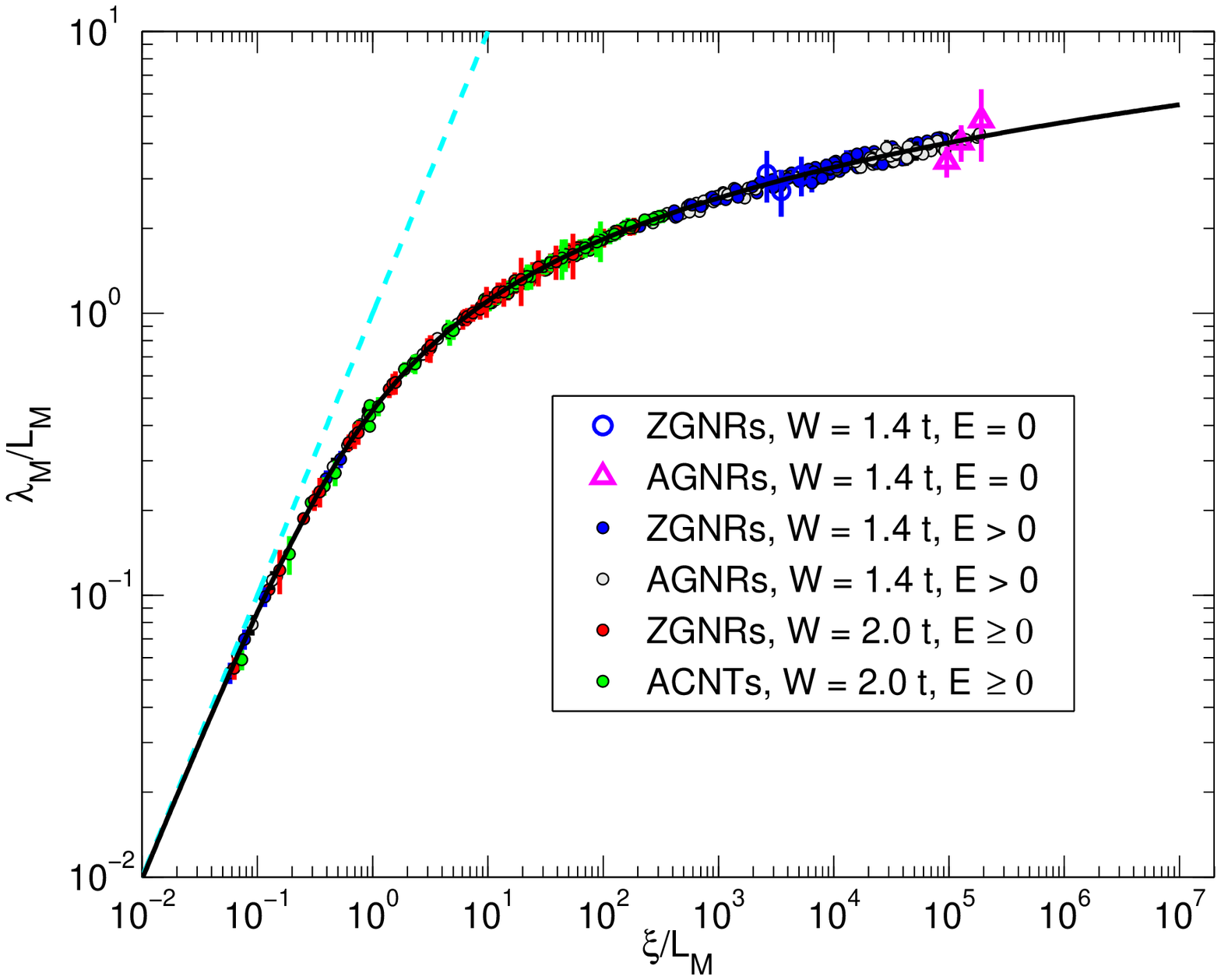}\\
  \caption{(color online) 
One-parameter scaling of localization length. The localization length divided by the width, 
$\lambda_M/L_M$, is plotted as a function of $\xi/L_M$, where $\xi$ is the 2D localization 
length obtained by fitting the data against the scaling curve. All the data from Fig.~\ref{figure:lambda_M} 
with the last three largest $M$ in each subfigure are considered. Abnormal data for the 
charge neutrality point in systems with weak disorder ($W=1.4t$) are emphasized. Due to the 
symmetry of the band structure, data with $E<0$ from Fig. 1 are omitted. The solid line 
represents the scaling function given by Eq.~(\ref{equation:scaling_function}) with $a=\pi$ 
and the dashed line represents the identity function $f(x)=x$. The error bars correspond to the
error estimates of $\lambda_M$ indicated in Fig.~\ref{figure:lambda_M}.}
\label{figure:sps_xi}
\end{center}
\end{figure*}

As our results indicate that the differences of localization lengths between
different boundary conditions and edge types become smaller with increasing width,
a natural question is whether the conventional one-parameter scaling theory of localization 
length applies to our simulation data. MacKinnon and Kramer 
\cite{mackinnon1983,mackinnon1981} have proposed the following scaling law 
for the Q1D localization length:
\begin{equation}
 \frac{\lambda_M}{L_M} = f\left(\frac{\xi}{L_M}\right),
\end{equation}
where $\xi=\xi(W,E)$ is the 2D localization length for a given $W$ and $E$, 
and $f=f(x)$ is an unknown function. The construction of the scaling function 
for graphene (or honeycomb lattice) has been considered by Schreiber and Ottomeier 
\cite{schreiber1992} as early as in 1992, although they only considered relatively 
strong disorder ($W\geq 4t$) due to the limited computational power available at that time. 
Recently, Lee \textit{et al.} \cite{lee2013} constructed a scaling curve for systems with 
$W$ down to $1.2t$, although not all the energy points (especially some points at and around 
the CNP) were considered uniformly. An inspection of the scaling curves presented in 
Refs. [\onlinecite{schreiber1992}] and [\onlinecite{lee2013}] reveals that the scaling 
function $f(x)$ may be universal. Thus, it is natural to attempt to construct an analytical 
expression for this scaling function.

To find such a universal function, we note that when $L_M$ is in the Q1D limit, 
where $L_M \ll \xi$ (i.e., $x \gg 1$) (but $L_M$ should be large enough to 
ensure that $\lambda_M/L_M$ enters the scaling regime), $\lambda_M/L_M$ 
decays nearly linearly with increasing $\ln (L_M)$ (not shown here). This 
indicates that $f(x) = a_1\ln (x)+a_2$, where $a_1$ and $a_2$ are constants. 
This kind of asymptotic behavior was in fact noticed very early by MacKinnon 
and Kramer \cite{mackinnon1983}. On the other hand, they also noted that when 
$L_M \gg \xi$ (i.e., $x \ll 1$), $\xi \approx \lambda_M$ and the scaling function 
should behave as $f(x) \sim x$. A natural choice for the scaling function which meets 
these two conditions simultaneously is thus $f(x) = \ln(1+k x)/k$, or equivalently,
\begin{equation}
\label{equation:scaling_function}
 \frac{\lambda_M}{L_M} = \frac{\ln \left(1 + k \xi/L_M\right)}{k},
\end{equation}
where $k$ is a constant which needs to be determined numerically. Before testing this 
function against our data, we point out that finding a parametrized analytical expression 
for the scaling function is not in sharp contrast with previous works. On the one hand, 
it is conventional to assume an analytical form for the scaling function when studying 
Anderson localization in three-dimensional systems \cite{slevin1999,rodriguez2010}, and 
following this approach, different functions have been tested for simulation data for 
graphene flakes \cite{gonzalez2013}. On the other hand, it has been assumed that in the 
limit of $x \ll 1$ the scaling function takes the following parametrized form \cite{lee2013, mackinnon1983}: 
\begin{equation}
\label{equation:expansion}
 f(x) = x - b x^2 + O(x^3),
\end{equation}
where $b$ is a fitting parameter. It is clear that Eq.~(\ref{equation:scaling_function}) 
automatically results in this kind of asymptotic behavior when $b=k/2$.

We have fitted the data of Fig.~\ref{figure:lambda_M} against Eq.~(\ref{equation:scaling_function}), 
treating the 2D localization lengths $\xi(E, W)$ for every $E$ and $W$ as independent fitting parameters. 
The results are shown in Fig.~\ref{figure:sps_xi}. We have only used the data for the 
three systems having the largest localization lengths in each of the Figs.~\ref{figure:lambda_M}(a)-(d)), 
since data for relatively narrow systems apparently do not follow any scaling curve. 
Nevertheless, our data already spread over a broader range of system widths compared 
to previous works \cite{schreiber1992,lee2013}. Accidentally or not, we estimate that 
the value of the parameter $k$ in Eq. ~(\ref{equation:scaling_function}) is very close to $\pi$. 
As can be seen from Fig.~\ref{figure:sps_xi}, all the data points project 
well onto the scaling curve, except for the CNP in the two weakly 
disordered ($W=1.4t$) systems. The reason why the CNP experiences the largest 
finite-size effect will be discussed later.
The scaling function, Eq.~(\ref{equation:scaling_function}) with $a=\pi$,
also gives an excellent description for the data in Ref.~[\onlinecite{lee2013}] \cite{note_private},  
as well as for the data for k square lattice with uncorrelated Anderson disorder, as shown in 
Appendix \ref{section:square}, and for the data for graphene with vacancy-type disorder, as will be 
discussed in Section \ref{section:vacancy_b}. While the simulation data
agree well with the proposed scaling function, in the next subsection
we will further explore its connection to another widely used method of computing the 2D localization length.

\subsection{Comparing two methods of computing the 2D localization length}

\begin{figure}
\begin{center}
  \includegraphics[width=\columnwidth]{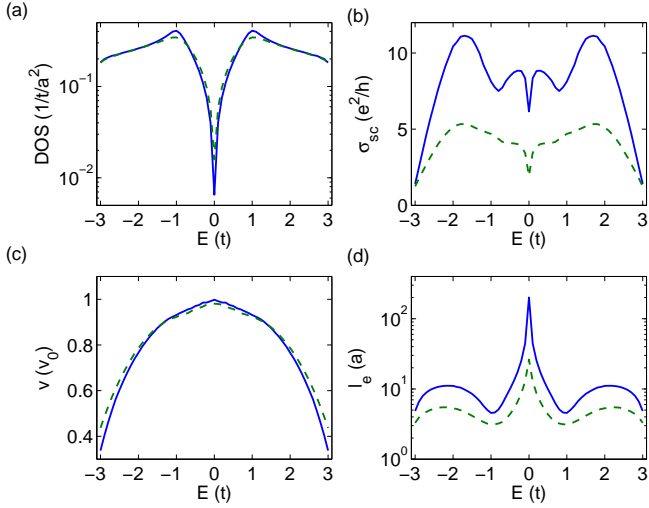}\\
  \caption{(color online) 
(a) Density of states, (b) semiclassical conductivity, (c) group velocity ($v_0=3at/2\hbar$), 
and (d) mean free path as functions of energy. The solid and dashed lines represent 
the results for $W=1.4t$ and $W=2.0t$, respectively. Sufficiently large simulation
cell sizes are used to eliminate the finite-size effects.}
\label{figure:diffusive_anderson}
\end{center}
\end{figure}

\begin{figure}
\begin{center}
  \includegraphics[width=\columnwidth]{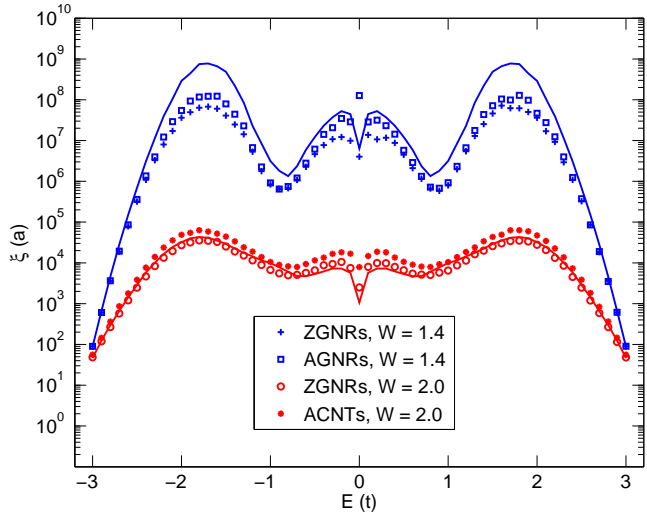}\\
  \caption{(color online)
Two-dimensional localization lengths as a function of energy. The markers are obtained by 
fitting the Q1D data (the same as used in Fig.~\ref{figure:sps_xi}) against 
Eq.~(\ref{equation:scaling_function}), with the specific types of the system indicated by 
the legends.  The lines are obtained by using Eq.~(\ref{equation:xi_from_sigma}), 
using the diffusive transport properties shown in Fig.~\ref{figure:diffusive_anderson}.}
\label{figure:xi_2d}
\end{center}
\end{figure}

According to the scaling theory of Anderson localization \cite{lee1985,sheng1995,rammer1998,dittrich1998}, 
$\xi$ can also be estimated exclusively based on the diffusive transport properties \cite{note_on__convention}:
\begin{equation}
 \label{equation:xi_from_sigma}
 \xi(E) = 2 l_\textmd{e}(E) \exp\left[\frac{\pi \sigma_{\textmd{sc}}(E)}{G_0}\right].
\end{equation}
It is thus important to ask whether this expression is consistent with the scaling approach 
based on the Q1D localization length. To answer this question, we first calculate the 
diffusive transport properties for systems with $W=1.4 t$ and $2.0t$. The results are shown in 
Fig.~\ref{figure:diffusive_anderson}. Note that the results are not sensitive to the edge type 
or boundary conditions, since the relevant transport length scale, the mean free path $l_{\textmd{e}}$, 
is relatively small (compared to $\xi$), and we can use a sufficiently large simulation cell size 
to eliminate any finite-size effects affecting the diffusive transport properties. 
An examination of Fig.~\ref{figure:diffusive_anderson} reveals why the CNP behaves very differently 
from other states regarding the localization properties. At the CNP, the density of states is 
vanishingly small but the semiclassical conductivity and the group velocity are of the same order 
as for other states. This results in a very large $l_{\textmd{e}}$ at the CNP, as has also been found
by Lherbier \textit{et al.} \cite{lherbier2008}.
With a disorder strength of $W=1.4t$, $l_{\textmd{e}} \approx 200 a$ at the CNP, 
which is comparable to the simulation widths used for calculating the Q1D localization lengths. 
One cannot expect that the scaling function applies when $L_M \sim l_{\textmd{e}}$, 
because $l_{\textmd{e}}$ sets up a lower limit of the scaling behavior \cite{mackinnon1983}. 
More quantitatively, $L_M$ should be at least several times larger than $l_{\textmd{e}}$ to make the 
scaling function fully applicable. However, with decreasing disorder strength, $l_{\textmd{e}}$ 
for the CNP diverges and it becomes formidable to reach the scaling regime computationally.

Figure~\ref{figure:xi_2d} compares the localization lengths calculated by 
Eq.~(\ref{equation:scaling_function}) (with $k=\pi$) and Eq.~(\ref{equation:xi_from_sigma}). 
We can see that the 2D localization lengths are much larger than the Q1D values, 
making a direct computation nearly impossible. They also depend sensitively on the 
disorder strength, with the values for $W=1.4 t$ being several orders of magnitude larger 
than those for $W=2.0t$. With a given disorder strength, the values of $\xi$ obtained using 
Eq.~(\ref{equation:scaling_function}) with different boundary conditions and edge types are 
very close to each other, only exhibiting some discrepancies around the CNP, which, as have 
been noted before, should be originated from the finite-size effect. It can be seen that the 
two methods for computing $\xi$ agree well with each other. Lee \textit{et al.} \cite{lee2013} 
also compared these two methods, but in contrast to our results, observed that 
Eq.~(\ref{equation:xi_from_sigma}) results in a significant underestimation. Our interpretation 
is that their method of computing $\sigma_{\textmd{sc}}$ is based on the semiclassical 
self-consistent Born approximation, which may be not as accurate as the fully 
quantum mechanical RSKG method.

The fact that Eq.~(\ref{equation:scaling_function}) and Eq.~(\ref{equation:xi_from_sigma}) 
give consistent results for $\xi$ can be understood in the following way. We know that in 
the Q1D limit, the localization length and the mean free path are related by the Thouless 
relation \cite{thouless1973,beenakker1997,avriller2007,uppstu2012} (for the orthogonal 
universality class, which is the case for graphene with intervalley scattering) \cite{note_on__convention}:
\begin{equation}
 \lambda_M(E) \approx N_{\textmd{c}}(E) l_{\textmd{e}}(E),
\end{equation}
where  $N_{\textmd{c}}(E)$ is the number of transport channels. In other words, $N_{\textmd{c}}(E)$ 
equals the ``hypothetical'' ballistic conductance as given by Eq.~(\ref{equation:ballistic_conductance}) 
divided by the conductance quantum $G_0\equiv2e^2/h$:
\begin{equation}
 N_c(E) \equiv \frac{g(E)}{G_0} = \frac{L_M e^2 \rho(E)v(E)}{2G_0}.
\end{equation}
By ``hypothetical'', we mean that $g(E)$ is the conductance of the disordered system in the zero length 
limit, where no scattering starts to play a role. By combining the above two equations and using the 
relation between $\sigma_{\textmd{sc}}(E)$ and $l_{\textmd{e}}(E)$ in Eq.~(\ref{equation:mfp}), 
we arrive at the following modified version of the Thouless relation:
\begin{equation}
 \lambda_M(E) = \frac{L_M \sigma_{\textmd{sc}}(E)}{G_0}.
\end{equation}
In the Q1D limit, the scaling function given by Eq.~(\ref{equation:scaling_function}) 
(with $k=\pi$) can be written as $\lambda_M(E)/L_M=\ln\left(\pi\xi(E)/L_M\right)/\pi$, which, 
combined with the above Thouless relation, gives
\begin{equation}
  \xi(E) = \frac{L_M}{\pi} \exp\left[\frac{\pi \sigma_{\textmd{sc}}(E)}{G_0}\right].
\end{equation}
Choosing $L_M=2\pi l_{\textmd{e}}(E)$ gives exactly Eq.~(\ref{equation:xi_from_sigma}). 
This heuristic derivation is consistent with the intuition that the scaling regime starts 
from a width several times larger than the mean free path.

\subsection{One-parameter scaling of conductivity}

\begin{figure}
\begin{center}
  \includegraphics[width=\columnwidth]{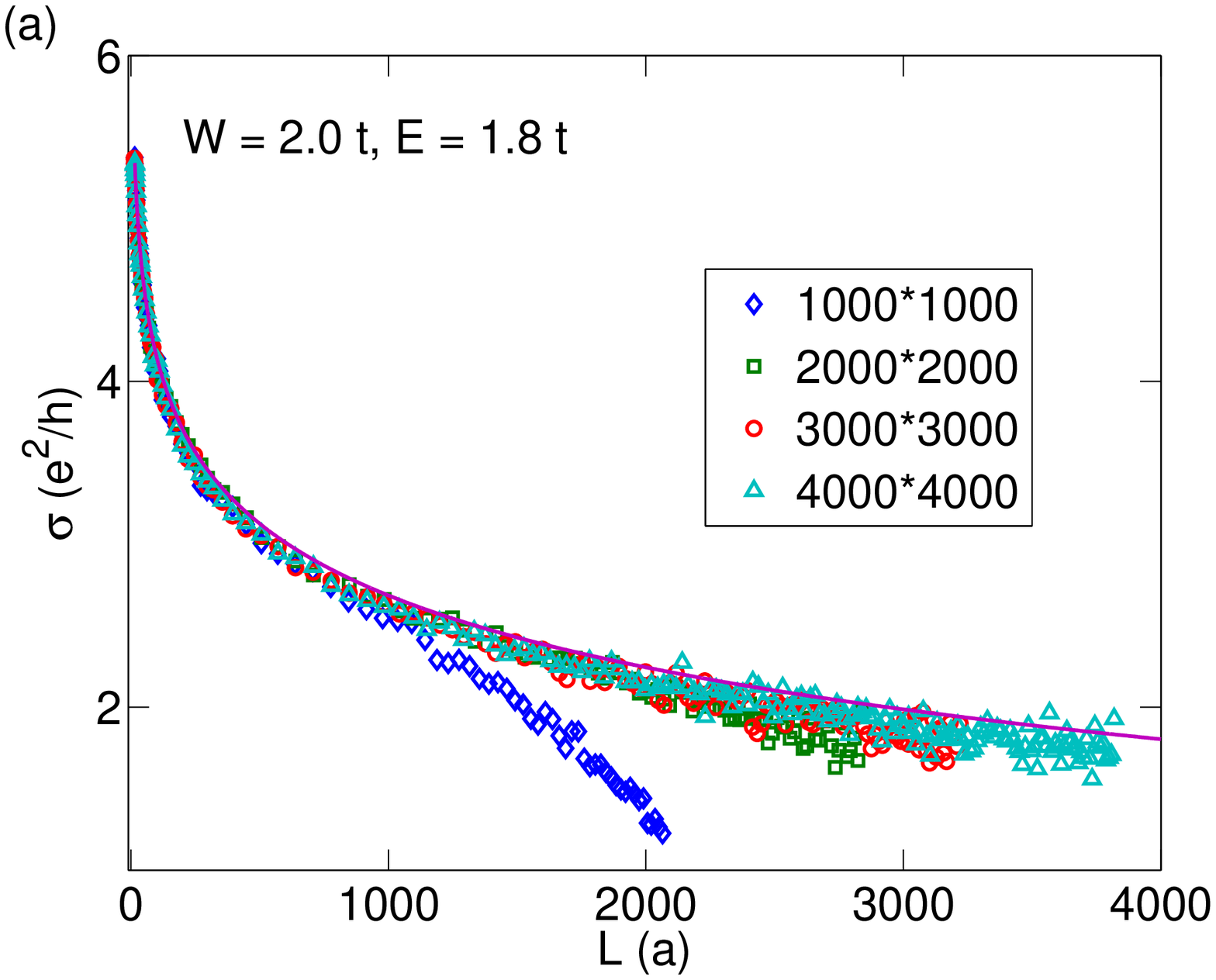}\\
  \includegraphics[width=\columnwidth]{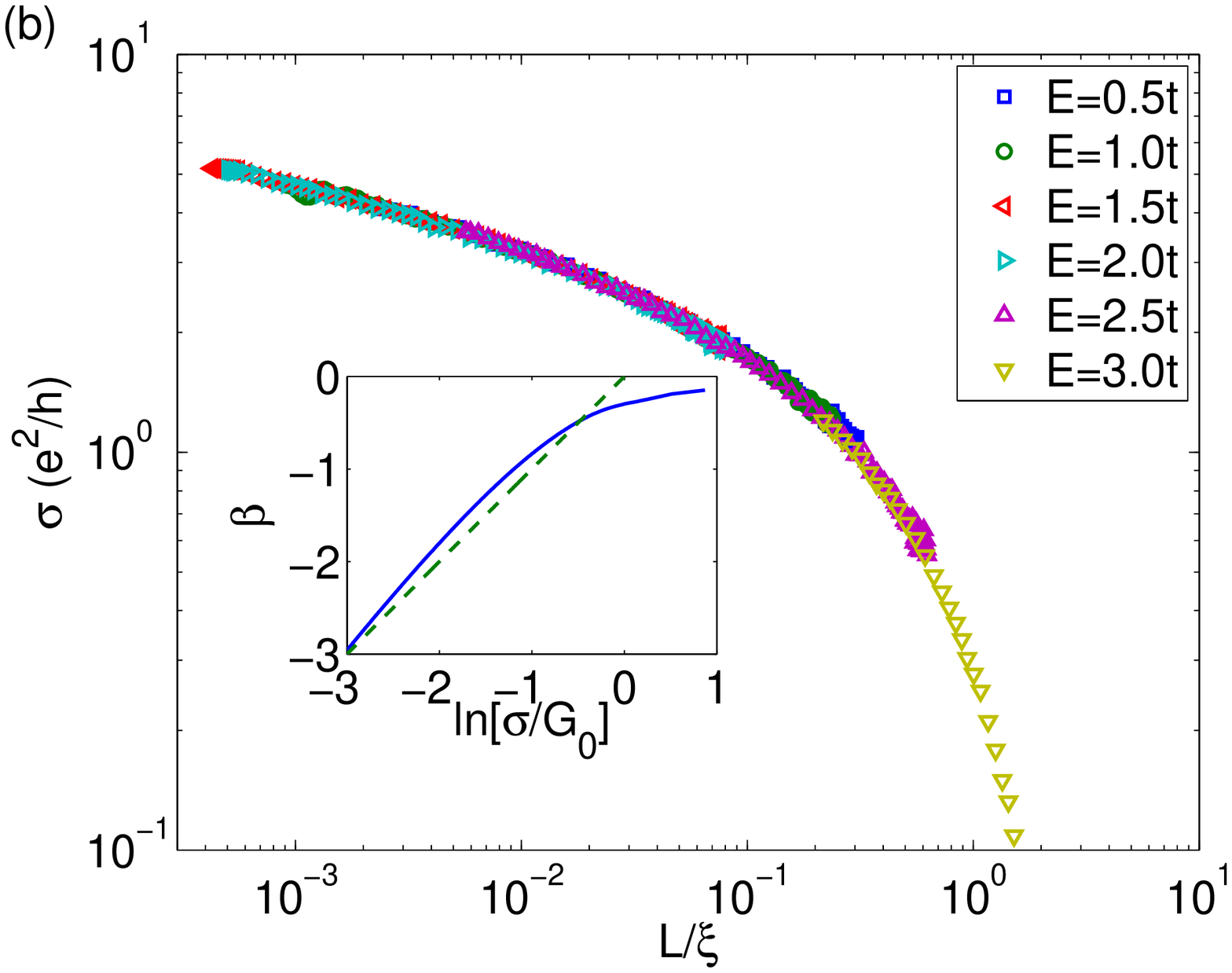}\\
  \caption{(color online)
Conductivity for 2D graphene with $W=2.0t$. (a) Conductivity as a function of the propagating 
length of electrons for different simulation sizes $N_x*N_y$ (markers). The prediction from 
the weak-localization formula given by Eq.~(\ref{equation:weak_localization}) is also shown (line). 
The energy considered  here is $E=1.8t$. (b) Conductivity as a function of the reduced length $L/\xi$ 
for a set of energy points. The 2D localization length $\xi$ is taken to be the average over the 
results obtained shown in Fig.~\ref{figure:xi_2d}. The inset in (b) shows the 
renormalization group $\beta$ function (solid line) calculated by using Eq.~(\ref{equation:beta}) after fitting $\sigma$ 
as a smooth function of $L/\xi$. The dashed line in the inset represents $\beta=\ln(\sigma/G_0)$.
Periodic boundary conditions are applied in both the transport 
and the transverse directions. The transport direction is taken to be along the zigzag edge;
taking the transport direction to be along the armchair edge yields similar results.  }
\label{figure:sps_sigma}
\end{center}
\end{figure}

The one-parameter scaling of localization length is in fact intimately connected \cite{mackinnon1983} 
to the one-parameter scaling of conductivity. 
Equation (\ref{equation:xi_from_sigma}) has been derived from the scaling behavior of the 2D 
conductivity in the  weak localization regime, where the conductivity $\sigma(E, L)$ decays 
logarithmically with increasing $L$:
\begin{equation}
 \label{equation:weak_localization}
 \sigma(E, L) = \sigma_{\textmd{sc}}(E) -
  \frac{G_0}{\pi} \ln\left[ \frac{L}{l_0(E)} \right].
\end{equation}
Here $l_0(E)$ is a length scale, conventionally set to $l_\textmd{e}(E)$. Assuming that $L$ 
reaches $\xi(E)$ when the weak localization correction becomes comparable to $\sigma_{\textmd{sc}}(E)$ 
gives Eq. (\ref{equation:xi_from_sigma}) apart from a factor of 2 resulting from the use of different 
conventions \cite{note_on__convention}.

The validity of the weak localization formula, Eq.~(\ref{equation:weak_localization}), 
can also be confirmed numerically.  Figure~\ref{figure:sps_sigma}(a) shows the calculated 
conductivity as a function of the propagating length, as defined by Eq.~(\ref{equation:length}), 
for the state with $E=1.8t$ and $W=2.0t$. The calculated conductivities are ensemble averaged over 
several disorder realizations and the tracing operation in Eq. (\ref{equation:msd}) has been 
approximated using several random vectors, resulting in relatively smooth curves. Due to the 
large localization length in 2D, significant finite-size effects arise when calculating the 
conductivity in the localized regime. When the simulation size $N_x \times N_y$ increases 
from $ 1000\times 1000$ to $4000 \times 4000$, the calculated data get closer to the line predicted by 
Eq.~(\ref{equation:weak_localization}), with $l_0(E)$ being set to the ``diffusion length'' 
$l_{\textmd{diff}}$ (which is generally larger than the mean free path) beyond which 
the conductivity starts to decay. $l_\text{diff}$ is defined as the length at which the running 
conductivity reaches its maximum value \cite{leconte2011,lherbier2012}. Although 
periodic boundary conditions are applied in both the transport and the transverse directions, 
we see that a simulation size of $1000\times 1000$ is not large enough to eliminate the 
finite-size effect, resulting in an artificial fast decay of conductivity when $L>1000a$.

The transition from the weak to the strong localization regime is smooth and universal. 
Figure~\ref{figure:sps_sigma}(b) shows the conductivity as a function of the propagating 
length normalized by the 2D localization length. The data for different energy states project 
onto a single curve, which agrees with the scaling theory of localization. This indicates the 
existence of a universal renormalization group $\beta$ function:
\begin{equation}
 \label{equation:beta}
 \beta = \frac{d \ln (\sigma/G_0)}{d \ln (L/\xi)},
\end{equation}
as shown in the inset of Fig.~\ref{figure:sps_sigma}(b). The scaling function behaves as 
$\beta\sim\ln\left(\sigma/G_0\right)$ when $\sigma\ll G_0$, which is consistent with the 
exponential decay of conductivity in the strongly localized regime. Similar results have 
been obtained \cite{bang2010} for hydrogenated graphene using the Landauer-B\"uttiker approach.
One may note that different renormalization group $\beta$ functions,
either with \cite{ostrovsky2007} or without \cite{bardarson2007} an unstable fixed 
point, have been obtained for graphene with long-range disorder.
While the positive sign of the $\beta$ functions (in the large conductivity limit)
in the previous works signifies antilocalization in the absence of intervalley scattering, 
the negative sign of the $\beta$ function in our 
work is associated with localization caused by intervalley scattering.

\section{\label{section:vacancy}Graphene with vacancy disorder}

Although the Anderson disorder model is of general theoretical interest, more realistic short-range 
scatterers in graphene are atomically sharp defects, such as vacancies and adatoms, 
which are believed to cause intervalley scattering and Anderson localization around the CNP in 
irradiated graphene\cite{chen2009} and hydrogenated graphene \cite{bostwick2009}. Here, we focus 
on the vacancy-type disorder, which  also approximates the effect of hydrogen adatoms \cite{wehling2010}.

\subsection{\label{section:vacancy_a}Finite-size effect resulting from the finiteness of the
simulation length}

\begin{figure}
\begin{center}
  \includegraphics[width=\columnwidth]{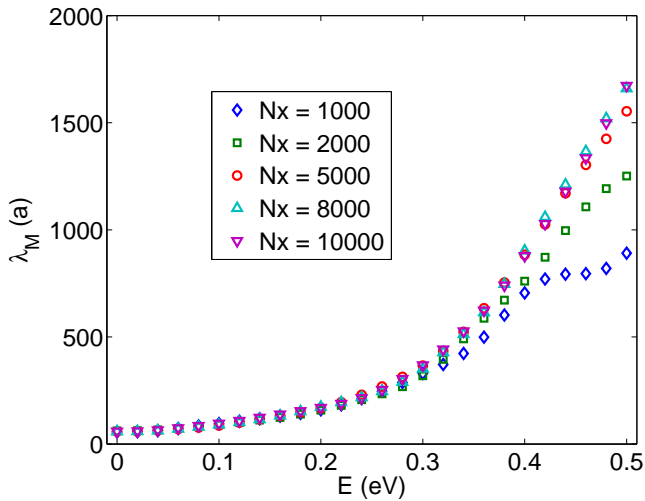}\\
  \caption{(color online) 
Demonstration of the finite-size effect for the calculation of the Q1D localization length
using the RSKG method. The Q1D localization length is plotted as function of energy. The
systems correspond to graphene (in the ACNTs geometry) with $1\%$ vacancies. The
width of the systems corresponds to a value of  $M=N_y=512$ (which gives $L_M=768a$)
and the simulation lengths are indicated by the $N_x$ (corresponding to a simulation cell
length of $\sqrt{3}N_x a / 2$) values in the legend. Error bars are
omitted, since their magnitudes are comparable to the marker size. }
\label{figure:size_effect_vacancy}
\end{center}
\end{figure}

Before presenting the results for graphene with vacancy defects, we first discuss 
the finite-size effect for the calculation of the Q1D localization length
using the RSKG method. This finite-size effect is different from that which causes the
deviations of the data for the CNP from the scaling function in Fig.~\ref{figure:sps_xi}.
It is a finite-size effect caused by the use of a finite simulation length in practical 
calculations. In the RSKG method, the propagating length $L(E, \tau)$, defined by Eq.~(\ref{equation:length}),
serves as a measure of the actual length of the physical system at a specific correlation time.
In contrast, the simulation cell length, which is proportional to $N_x$ (or $N_y$, depending on
the transport direction) has no direct connection to $L(E, \tau)$. Usually, periodic boundary
conditions are applied along the transport direction to alleviate the finite-size effect caused
by the finiteness of $N_x$. Whether or not a given $N_x$ is large enough to eliminate the
finite-size effect depends on the involved transport length scales. Figure~\ref{figure:size_effect_vacancy}
shows the finite-size effect when calculating the Q1D localization lengths for ACNTs of width $L_M=768a$
with $1\%$ vacancies. As the simulation cell length increases from $N_x=10^3$ to $N_x=10^4$, 
the calculated Q1D localization lengths converge, which reflects the alleviation of the 
finite-size effect by increasing the simulation cell length. It is clear to see that states with larger
saturated localization lengths require larger simulation cell lengths to eliminate the finite-size effect. 
More quantitatively, to completely eliminate the finite-size effect, the simulation cell length 
should be a few times larger than the maximum localization length for a given simulated system. 
In this paper, we have used as large as possible simulation cell lengths,
and the finite-size effects resulting from the finiteness of $N_x$ have been practically eliminated.

\subsection{\label{section:vacancy_b}One-parameter scaling of localization length}

\begin{figure*}
\begin{center}
  \includegraphics[width=16cm]{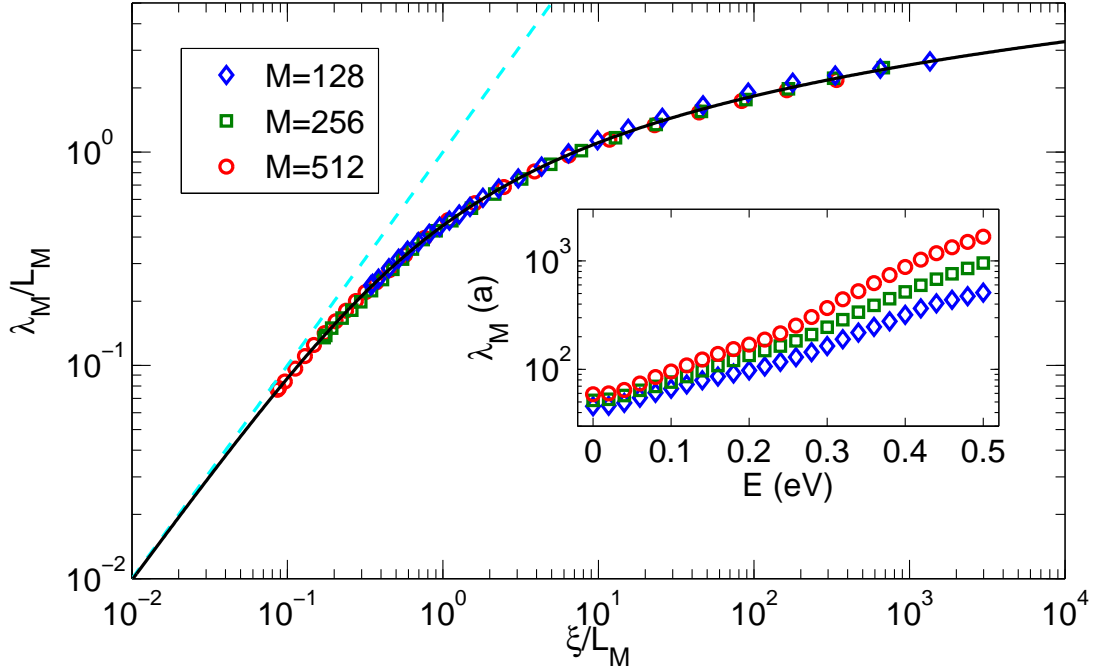}\\
  \caption{(color online) 
One-parameter scaling of localization length for graphene with $1\%$ vacancy disorder. 
The localization length divided by the width, $\lambda_M/L_M$, is plotted as a function 
of $\xi/L_M$, where $\xi$ is the 2D localization length obtained by fitting the data in the inset
against the scaling function. The solid line represents the scaling function given by 
Eq.~(\ref{equation:scaling_function}) with $k=\pi$ and the dashed line represents the identity 
function $f(x)=x$. The inset shows the Q1D localization lengths as a function of energy. 
The transport direction is along the zigzag edge and periodic boundary conditions are applied 
along the transverse direction for the Q1D systems. The Q1D systems have a fixed 
vacancy concentration of $1\%$.}
\label{figure:vacancy1}
\end{center}
\end{figure*}

We have calculated the localization lengths for Q1D graphene systems in the ACNT geometry
with $M=128$, 256, and 512, with the vacancy concentration fixed to $n=1\%$.
The results are shown in the inset of Fig.~\ref{figure:vacancy1}.
The main frame of Fig.~\ref{figure:vacancy1} shows that the scaling 
function given by Eq.~(\ref{equation:scaling_function}), 
with $k \approx \pi$, also applies here.  A striking difference 
between vacancy disorder and Anderson disorder is that the Van Hove singularities at $E=\pm t$ are much 
more strongly affected by Anderson disorder (manifested in the local minimum of the mean free path 
at $E=\pm t$ in Fig.~\ref{figure:diffusive_anderson}), while vacancies mostly affect low-energy 
charge carriers around the CNP. This is because vacancies serve as high potential barriers which 
result in large scattering cross sections and small mean free paths for low-energy 
charge carriers \cite{uppstu2012}. In contrast, high-energy charge carriers experience small 
scattering cross sections and have large mean free paths, which combined with higher densities of states 
(larger number of transport channels), gives rise to large Q1D localization lengths according to the Thouless relation. 
For the selected defect concentration, our numerical calculations are only able to explore a small 
energy range $|E|\leq 0.5$ eV around the CNP. Within this energy range, all the data agree well 
with Eq.~(\ref{equation:scaling_function}), and the corresponding 2D localization length can thus be extracted.

\subsection{\label{section:vacancy_b}Connecting diffusive and localized transport regimes}

\begin{figure}
\begin{center}
  \includegraphics[width=\columnwidth]{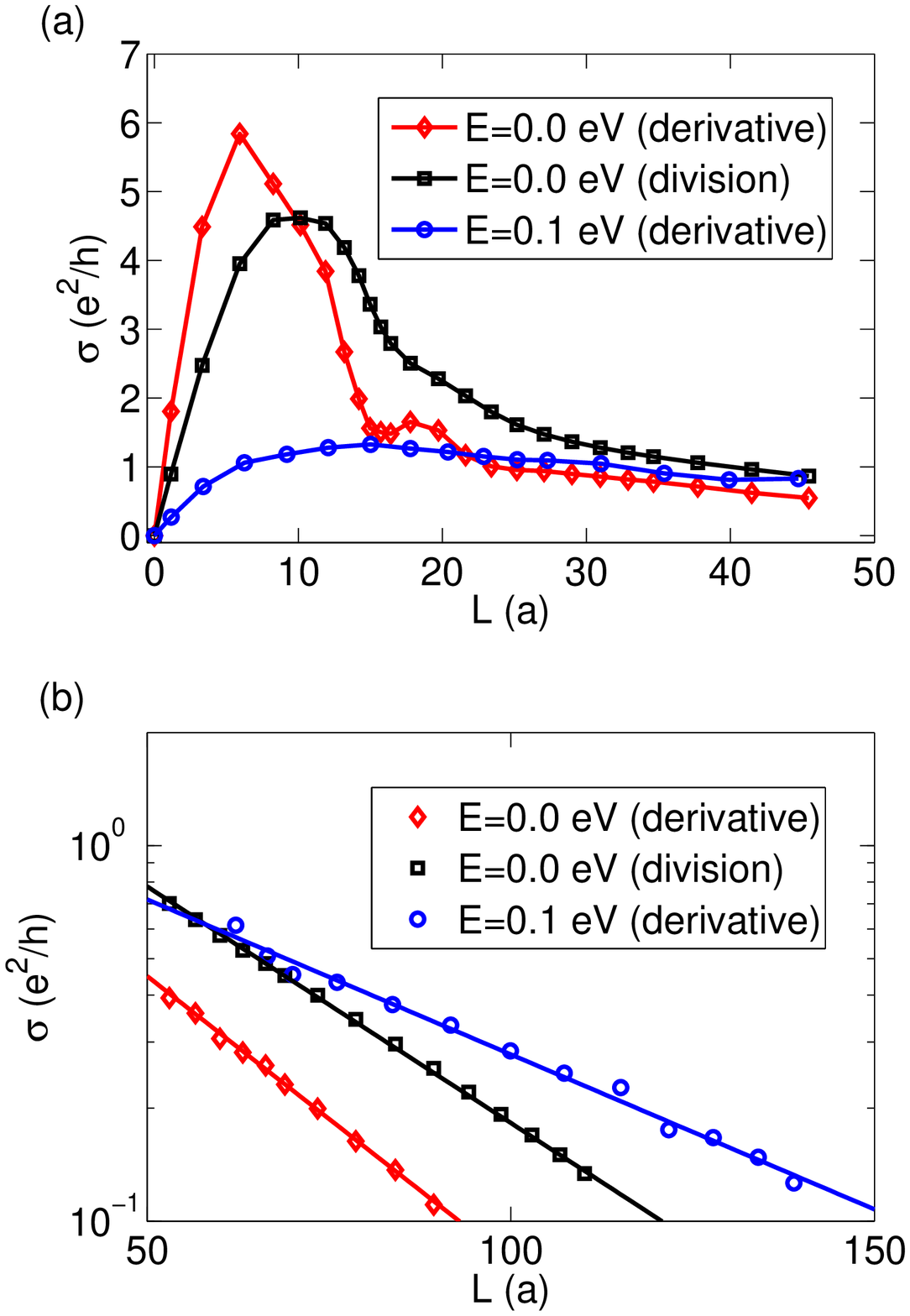}\\
  \caption{(color online) 
Conductivity as a function of propagating length in (a) the ballistic-to-diffusive transition
regime and (b) the localized regime. ``derivative'' in the legend means that the data
are obtained by using the derivative-based definition of the running conductivity, 
as given by Eq.~(\ref{equation:rec}), while ``division'' means that the data
are obtained by  substituting the time derivative  with a time division. The markers and 
lines in (b) represent raw data and exponential fits using $\sigma(L)\sim\exp(-2L/\xi)$,
respectively. The simulated system corresponds to 2D graphene (using a sufficiently large simulation 
cell size) with a vacancy concentration of $1\%$. }
\label{figure:vacancy_diffusive}
\end{center}
\end{figure}

\begin{figure}
\begin{center}
  \includegraphics[width=\columnwidth]{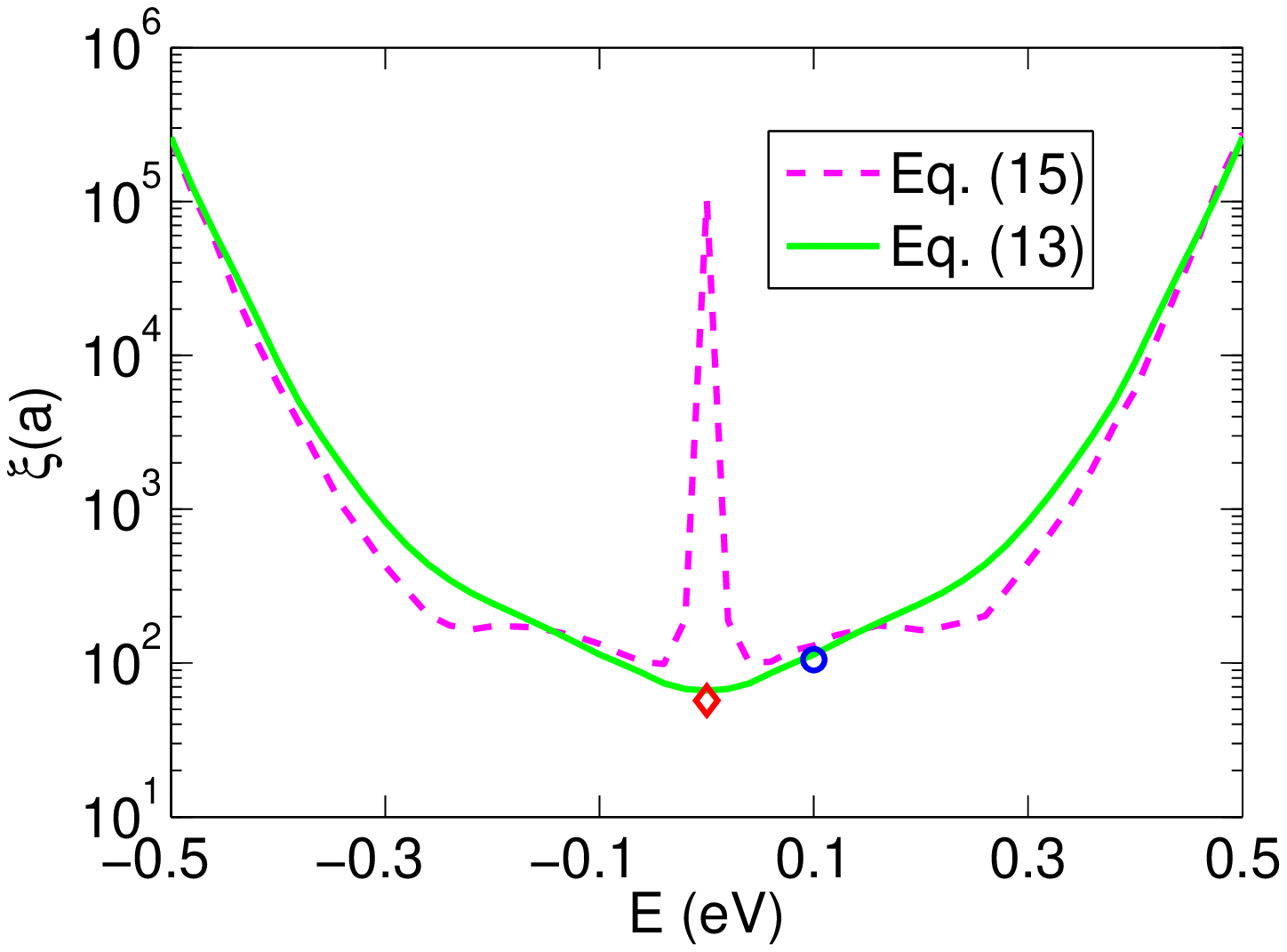}\\
  \caption{(color online) 
2D Localization length as a function of energy obtained by using Eq.~(\ref{equation:xi_from_sigma}) 
(dashed line) and Eq.~(\ref{equation:scaling_function}) (solid line). When using Eq.~(\ref{equation:xi_from_sigma}),
a sufficiently large simulation cell size is used to obtain the diffusive transport properties. When using
Eq.~(\ref{equation:scaling_function}), Q1D localization length data from the inset of 
Fig.~\ref{figure:vacancy1} are used. The diamond and circle correspond to the results obtained
by the exponential fitting as shown in Fig.~\ref{figure:vacancy_diffusive}(b) for the CNP (using the 
derivative-based definition for the running conductivity) and $E=0.1$ eV, respectively. 
The studied system corresponds to 2D graphene 
with a vacancy concentration of $1\%$.}
\label{figure:vacancy_compare}
\end{center}
\end{figure}

As in the case of graphene with Anderson disorder, one may ask whether the 2D localization lengths obtained 
by fitting the Q1D data against Eq.~(\ref{equation:scaling_function}) are consistent with those 
obtained by using  Eq.~(\ref{equation:xi_from_sigma}). It turns out that there is some ambiguity 
in the calculation of the semiclassical conductivity at the CNP, as shown in 
Fig.~\ref{figure:vacancy_diffusive}(a), where the running conductivity obtained by using Eq.~(\ref{equation:rec}) 
is compared with that obtained by substituting the time derivative in Eq.~(\ref{equation:rec}) 
with a time division. The latter may be well described by a power-law length-dependence
in an appropriate regime \cite{cresti2013,laissardiere2013}, 
and is thus associated with an infinite localization length,
as suggested in the previous works.
However, the correct derivative-based definition of $\sigma$ does not support the 
power-law length dependence. The calculated $\sigma(L)$ develops more than one peak, which may just reflect
the radial distribution profile  of the local density of states, which has large magnitude
in the vicinity of the vacancies \cite{ugeda2010}. In the RSKG method, as the wavepackets
(associated with individual sites) propagate, they can ``feel'' a large local density of
states associated with the conductivity peak before reaching the diffusive regime. 
Unfortunately, there does not seem to be any completely 
unambiguous method in the RSKG formalism for determining a diffusive regime where a well defined value 
of $\sigma_{\textmd{sc}}(E)$ can be extracted. When moving away from the CNP, the effect of the 
local density of states diminishes, and there is no such local peaks of conductivity, as
shown by the results for $E=0.1$ eV in Fig.~\ref{figure:vacancy_diffusive}(a).

The large local density of states at the CNP affects the conductivity significantly only in the 
ballistic-to-diffusive regime. In the strongly localized regime, we expect that the conductivity decays
exponentially with increasing length. This is confirmed by the results shown in 
Fig.~\ref{figure:vacancy_diffusive}(b). Here, the simulation data can be well described by 
the exponential fitting\cite{note_on__convention}: 
$\sigma(L)\sim\exp(-2L/\xi)$. Even the conductivity at the CNP 
obtained by approximating the time-derivative with a time-division  follows 
the exponential law in the strongly localized regime, although this approximation 
results in a much larger value of conductivity at a given length.

Figure~\ref{figure:vacancy_compare} shows the 2D localization lengths calculated by 
Eq.~(\ref{equation:xi_from_sigma}) and Eq.~(\ref{equation:scaling_function}), along with
those for $E=0$ and 0.1~eV extracted using the exponential fitting. Here, the 
semiclassical conductivity is taken to be the maximum of the running conductivity
when applying Eq.~(\ref{equation:xi_from_sigma}).
The agreement between Eq.~(\ref{equation:xi_from_sigma}) and Eq.~(\ref{equation:scaling_function})
is good only at higher energies. At the CNP, the prediction of Eq.~(\ref{equation:xi_from_sigma})
is far too large compared to that given by Eq.~(\ref{equation:scaling_function}). In contrast,
the exponential fitting gives rise to results consistent with Eq.~(\ref{equation:scaling_function}).
We thus conclude that the discrepancy between Eq.~(\ref{equation:xi_from_sigma}) 
and Eq.~(\ref{equation:scaling_function}) is largely resulted from the 
ambiguity in the calculation of the semiclassical conductivity.

\subsection{\label{section:resolution} Effects of energy resolution and vacancy concentration}

Due to the large density of states around the CNP, one may expect that the energy resolution $\delta E$
used in the numerical calculations would affect the results. 
To see how the energy resolution affects the results, we first calculate the density of states
and running conductivity for graphene with $1\%$ vacancy defects using different values of 
$N_m$, the number of Chebyshev moments in the kernel polynomial method. Although there
may be no exact relationship between $\delta E$ and $N_m$, it is generally believed \cite{weisse2006}
that $\delta E \propto 1/N_m$. Therefore, one can increase the energy resolution, i.e., decrease $\delta E$,
by increasing $N_m$. 

\begin{figure*}
\begin{center}
  \includegraphics[width=.66\columnwidth]{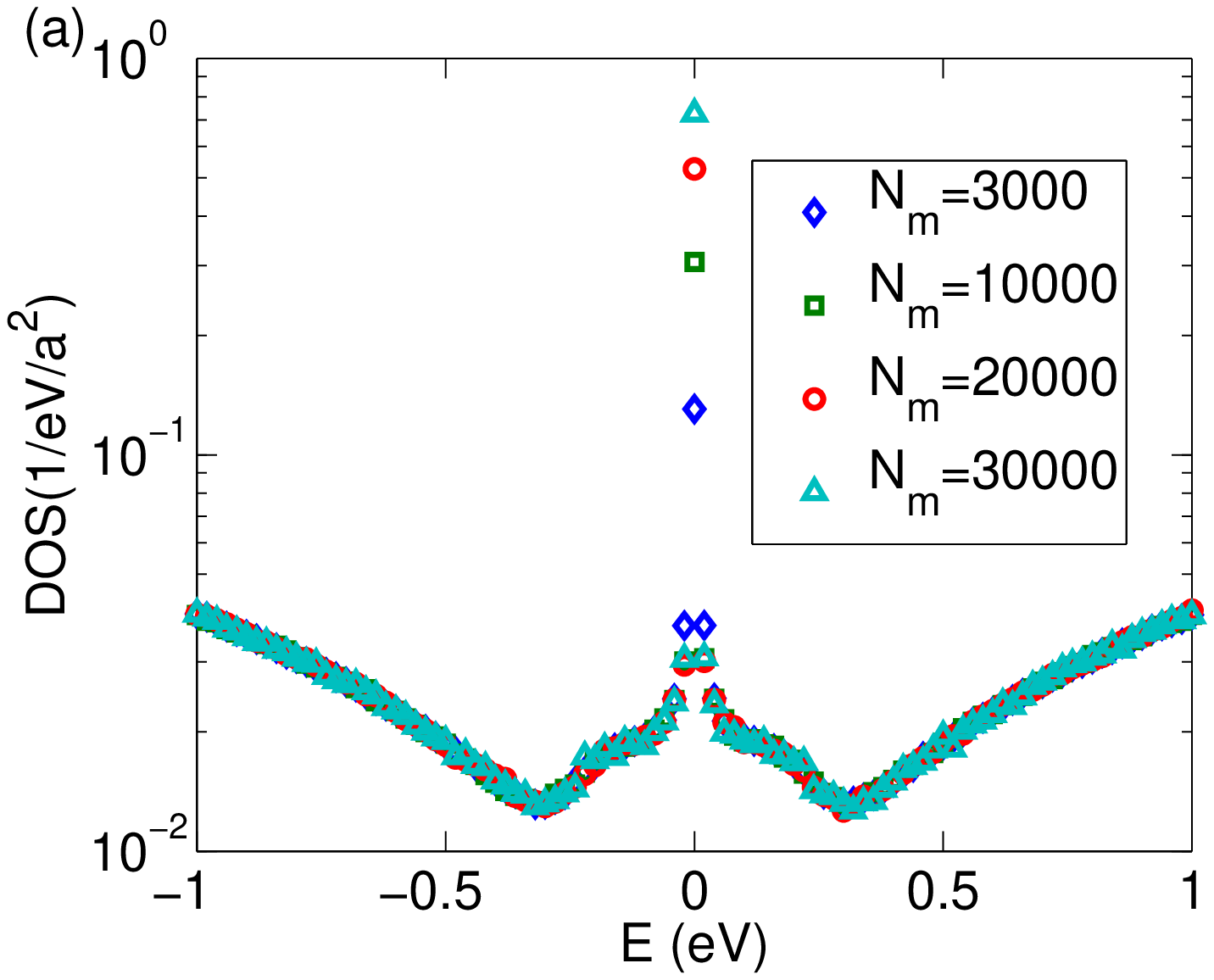}
  \includegraphics[width=.66\columnwidth]{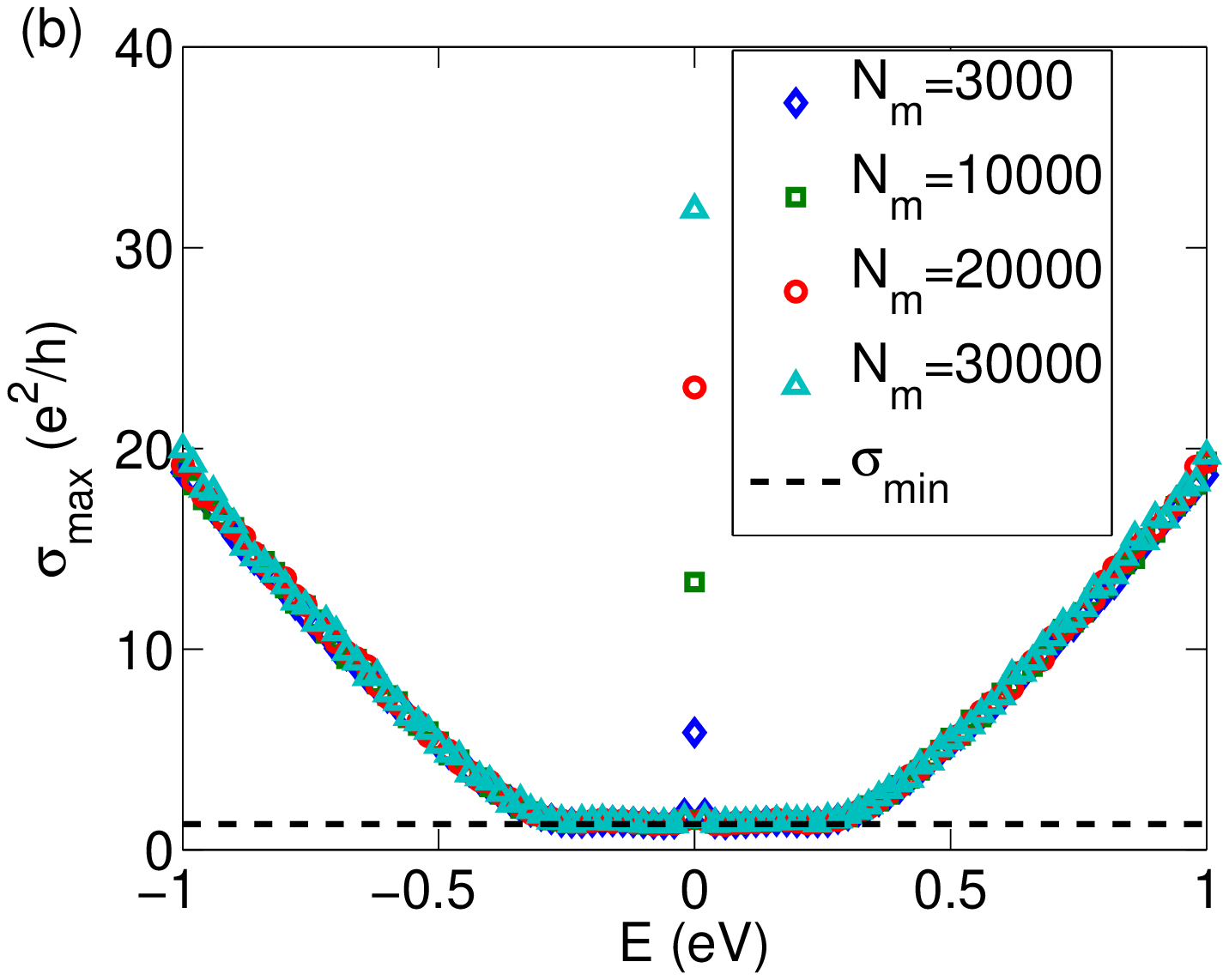}
  \includegraphics[width=.66\columnwidth]{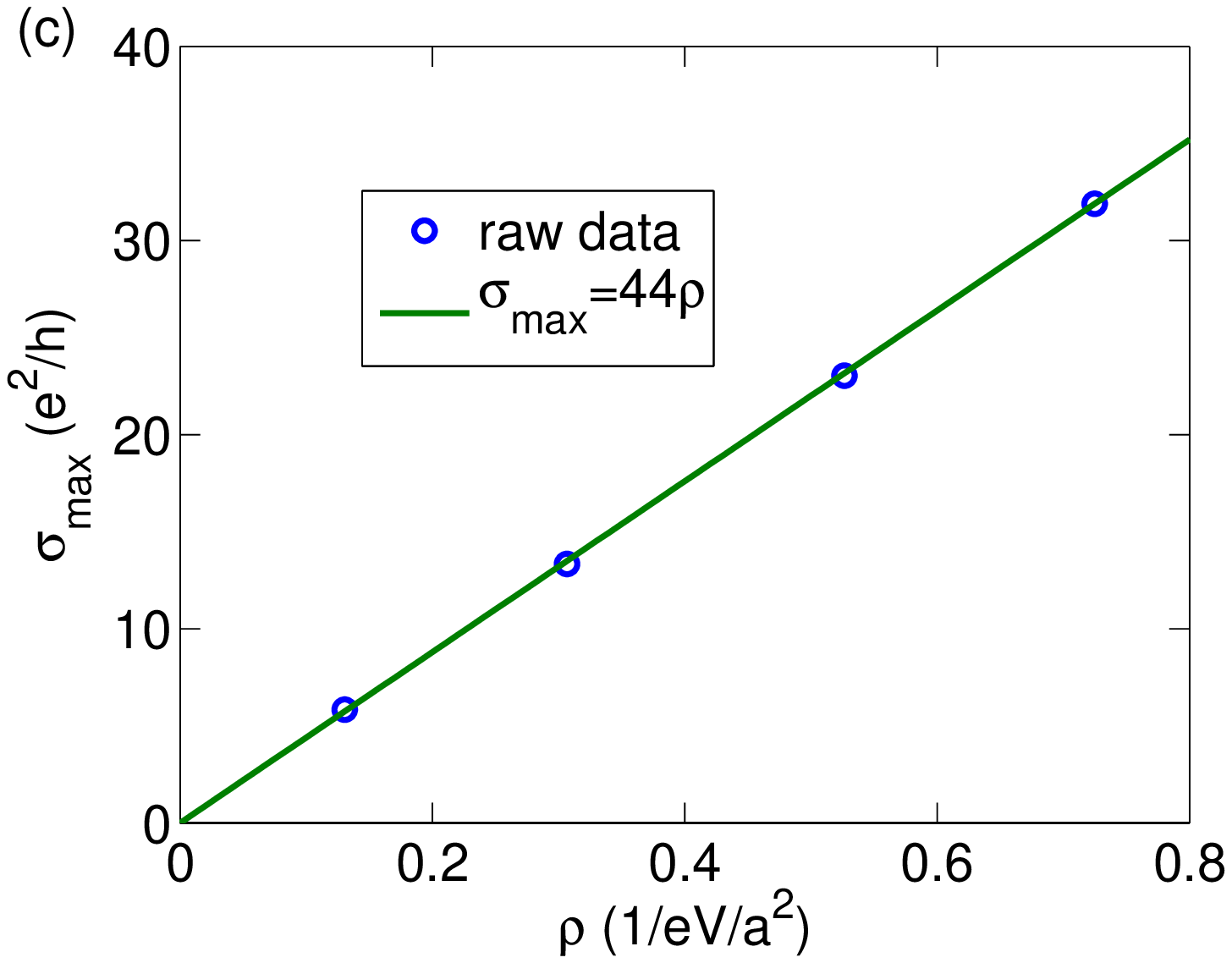}
  \caption{(color online) (a) Density of states and (b) maximum conductivity (over 
   correlation time) as a function of energy for 2D graphene with $1\%$ vacancy defects
   calculated by using different energy resolutions corresponding to
   different numbers of Chebyshev moments ($N_m$) used in the kernel polynomial 
   method. The dashed line in (b) indicates the ``minimum conductivity'' 
   $\sigma_{\text{min}} = 4e^2/(\pi h)$. (c) Maximum conductivity at the CNP
   as a function of the density of states $\rho$ at the CNP. The line in (c) represents 
   the linear dependence $\sigma_{\text{max}}(E=0)=44\rho(E=0)$. 
   To achieve high statistical accuracy, 
   $N_r=50$ random vectors were used for each energy resolution. 
 }
\label{figure:resolution_1}
\end{center}
\end{figure*}

\begin{figure}
\begin{center}
  \includegraphics[width=\columnwidth]{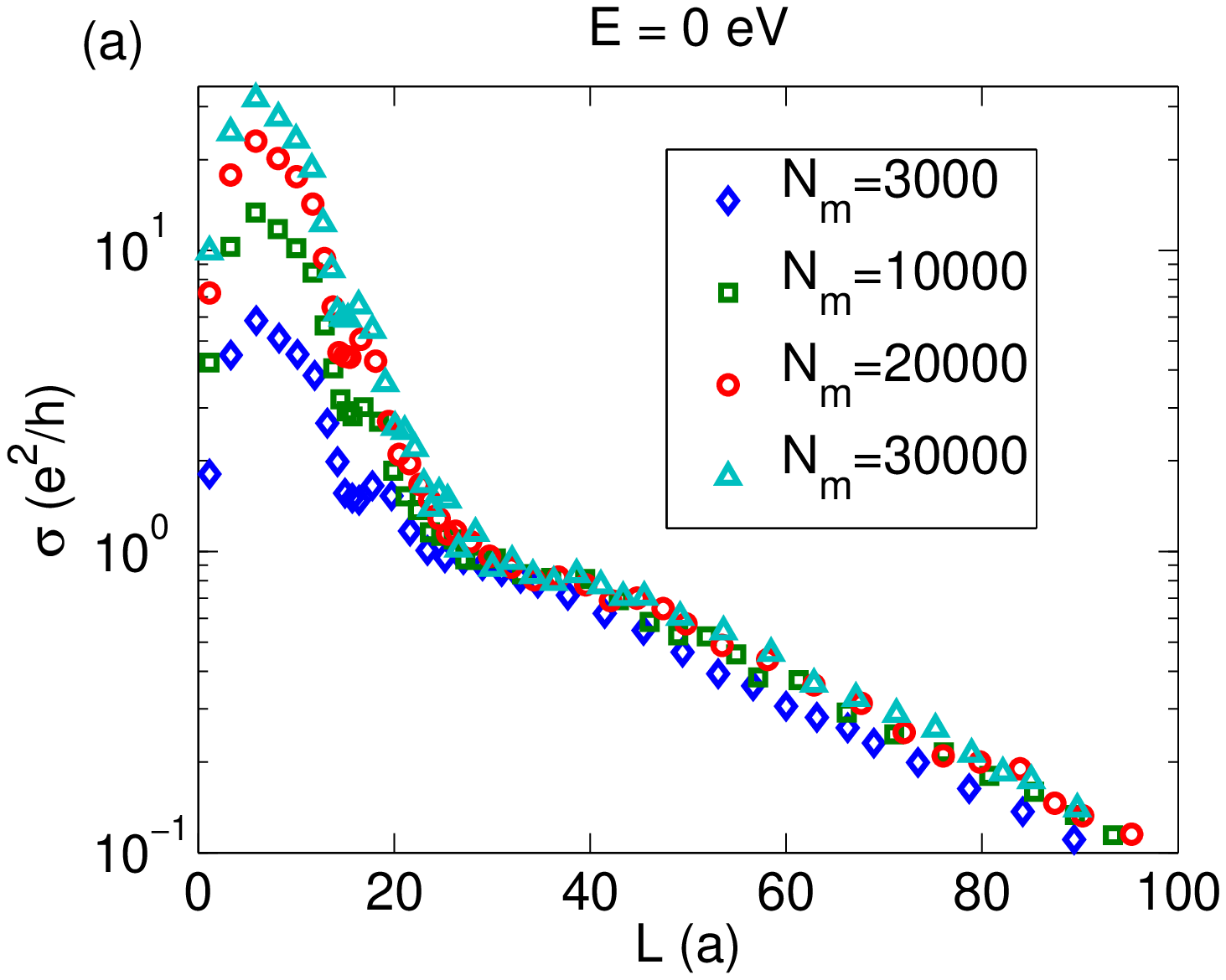}\\
  \includegraphics[width=\columnwidth]{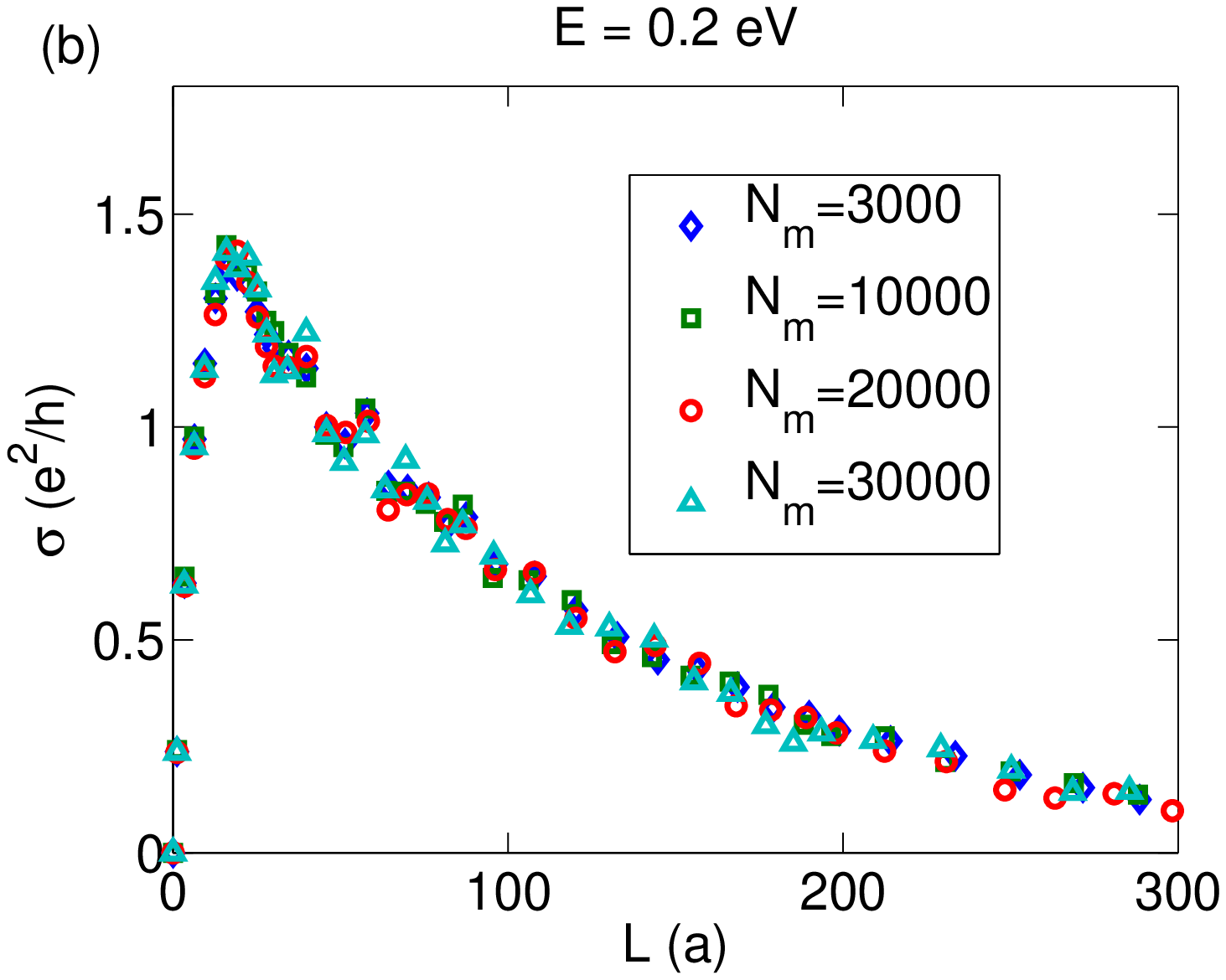}
  \caption{(color online)
   Running conductivity as a function of propagating length for (a) the CNP
   and (b) $E=0.2$ eV in 2D graphene with $1\%$ vacancy defects
   calculated by using different energy resolutions corresponding to
   different numbers of Chebyshev moments ($N_m$) used in the kernel polynomial 
   method. To achieve high statistical accuracy, 
   $N_r=50$ random vectors are used for each energy resolution. 
 }
\label{figure:resolution_2}
\end{center}
\end{figure}

Figure \ref{figure:resolution_1} presents the results for the density of states $\rho(E)$ and the maximum
conductivity $\sigma_{\text{max}}(E)$ (over the correlation time), the latter being conventionally 
taken as the definition of $\sigma_{\text{sc}}(E)$ in the RSKG method. It can be seen that with
increasing energy resolution, both $\rho(E)$ and $\sigma_{\text{max}}(E)$ develop
increasingly high values at the CNP. In contrast, the results for the other energy points do not
depend on the energy resolution. Interestingly, $\sigma_{\text{max}}(E=0)$ is proportional to
$\rho(E=0)$, as shown in Fig. \ref{figure:resolution_1}(c). Then, one may ask if the length-dependence
of the conductivity at the CNP also depends crucially on the energy resolution. To answer 
this question, we have plotted the running conductivity as a function
of the propagating length $L$ at the CNP, obtained by using different 
energy resolutions, in Fig. \ref{figure:resolution_2}(a). 
It can be seen that when $L<30a$, i.e., roughly in the ballistic-to-diffusive regime, 
the results depend strongly on the energy resolution.
Outside this regime, the dependence disappears
with increasing $N_m$, with the results being
converged when $N_m>10000$. Moreover, it can be seen that the energy resolution 
does not affect the obtained localization length.
Figure \ref{figure:resolution_2}(b) shows the running conductivity at $E=0.2$ eV, 
also obtained using different energy resolutions. The
energy resolution does not seem to significantly affect 
the results at any length scale away from the CNP.

\begin{figure}
\begin{center}
  \includegraphics[width=\columnwidth]{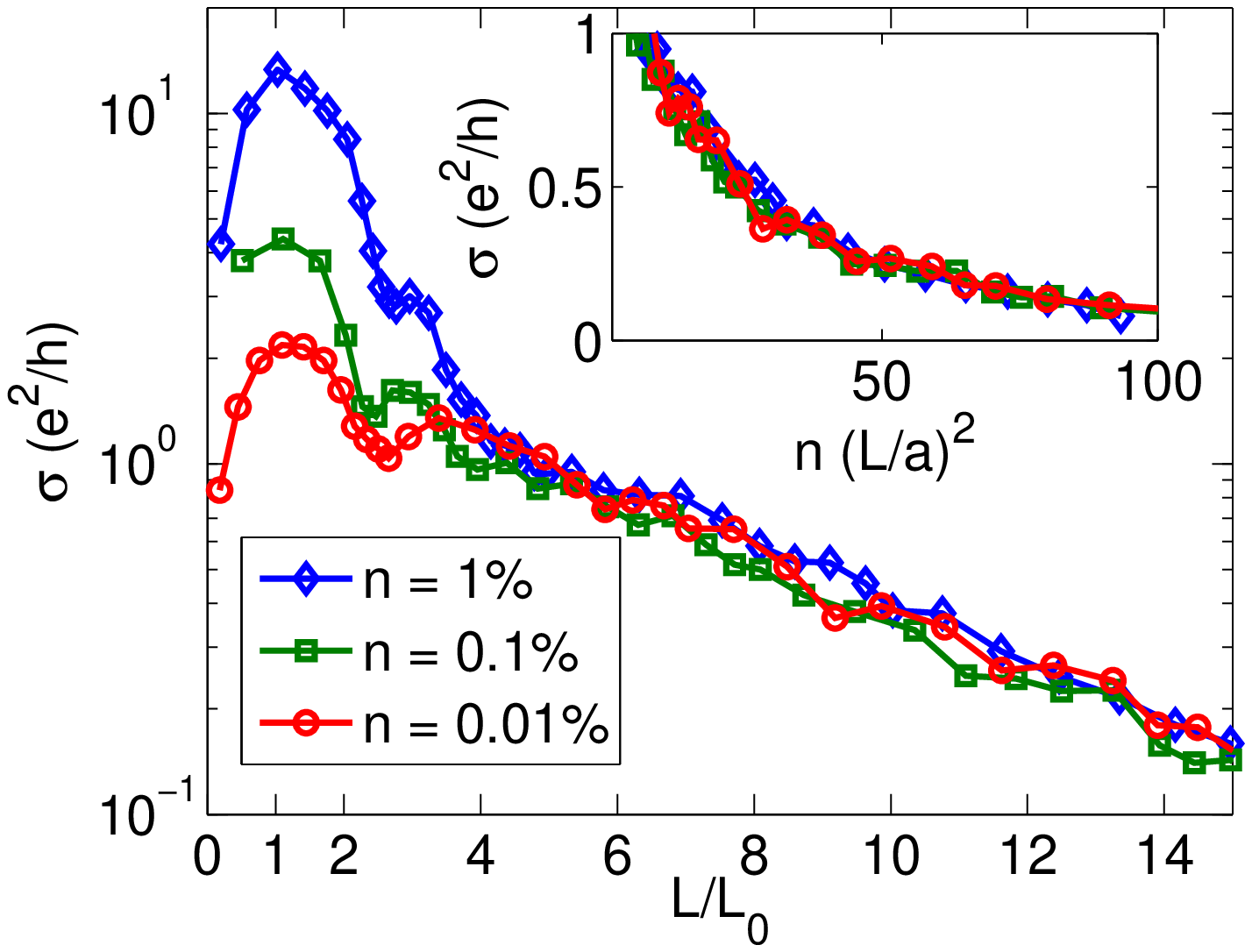}\\
  \caption{(color online) 
Running conductivity at the CNP as a function of the normalized propagating length $L/L_0$ in graphene
with vacancy defects, where
$L_0$ is the average distance between an atom and its nearest vacancy. The inset
shows the running conductivity as a function of $n(L/a)^2$ in the scaling regime, where $n$ is the 
vacancy concentration, as indicated in the legend. For all the vacancy concentrations,
the number of Chebyshev moments the number of random
vectors are chosen to be $N_m=10000$ and $N_r=50$, respectively. }
\label{figure:vacancy_different_n}
\end{center}
\end{figure}

So far, we have only considered a relatively large vacancy concentration of $n=1\%$. 
We now study how the defect concentration affects the scaling of conductivity at the CNP,
by additionally considering systems with lower vacancy concentrations: $n=0.1\%$ and $n=0.01\%$.
The results are shown in Fig. \ref{figure:vacancy_different_n}. 
In the main frame, we have plotted the running conductivity as a function of
the normalized propagating length $L/L_0$, where $L_0$ is the average distance
between an atom and its nearest vacancy. From simple geometric considerations, 
one can find that
\begin{equation}
 L_0 = \frac{1}{4} \sqrt{  \frac{3\sqrt{3}} {n}  } a,
\end{equation}
which can also be confirmed by numerical calculations. 
One can make several observations based on Fig. \ref{figure:vacancy_different_n}:

(1) The maximum values $\sigma_{\text{max}}$ of the running conductivity are different for different vacancy 
concentrations $n$; a higher $n$ gives a higher $\sigma_{\text{max}}$. This indicates that the peak of 
the running conductivity is related to the local density of states around the vacancies.
 
(2) For all the considered vacancy concentrations, the running conductivity takes its maximum
at $L=L_0$ ($L/L_0=1$ in Fig. \ref{figure:vacancy_different_n}). This further supports our
suggestion that the peak of the running conductivity is directly related to the local density of 
states around the vacancies, since $L_0$ is also the distance at which the radial distribution 
function of the local density of states attains its peak value.

(3) Beyond the ballistic-to-diffusive regime, i.e., when $\sigma<e^2/h$, the running conductivities
for different vacancy concentrations are well correlated and decay exponentially with
increasing length. This is strong evidence for the validity of the one-parameter scaling. 
Since $L_0 \propto n^{-1/2}$, the running conductivities are also correlated when
plotted as a function of $n(L/a)^2$, as shown in the inset of Fig. \ref{figure:vacancy_different_n}.
Our results are qualitatively different from those by Ostrovsky \textit{et al.} \cite{ostrovsky2010}. 
Using a different numerical method, they found that the running conductivity saturates to
a constant on the order of $\sigma_{\text{min}}$ with increasing $n(L/a)^2$, without 
localization even up to $n(L/a)^2 = 300$. We are not sure about the origin of the different
results, but we note that Ostrovsky \textit{et al.} have remarked that \cite{ostrovsky2010} 
the systems will eventually 
enter the localized regime with increasing vacancy concentration.

(4) Based on the correlation in the main frame of Fig. \ref{figure:vacancy_different_n}, we can infer
that the localization length is proportional to $L_0$, which is in turn proportional to 
the average distance between the vacancies. Based on the analysis of the effective cross sections \cite{uppstu2012},
we know that the mean free path is also proportional to $L_0$. Therefore, the (2D) localization
length at the CNP is directly proportional to the mean free path, indicating [according to 
Eq.~(\ref{equation:xi_from_sigma})] that $\sigma_{\text{sc}}$ at the CNP does 
not depend on the vacancy concentration. Taking the mean free path as $L_0$, we estimate that
$\sigma_{\text{sc}} \approx e^2/h$ at the CNP. Using this value for $\sigma_{\text{sc}}$,
the discrepancy between Eq.~(\ref{equation:xi_from_sigma}) and Eq.~(\ref{equation:scaling_function})
at the CNP disappears.

Although the CNP has a very large density of states coming from the resonant states (mid-gap states), 
it is the most localized state, exhibiting the smallest localization length. 
The state at the CNP is a quasilocalized state \cite{ugeda2010} 
and also exhibits a peak value of the inverse participation ratio \cite{pereira2006}. Therefore, 
Anderson localization can be observed around the CNP, manifesting itself as conductivities 
smaller than the minimum conductivity $\sigma_{\textmd{min}} = 2G_0/\pi$ of pristine graphene. 
However, when moving away from the CNP, the localization length increases quickly, even up to 
values much larger than realistic sample sizes or coherence lengths. For a fixed sample size, 
the localization effect is only significant around the CNP and disappears rapidly with 
increasing energy (or carrier concentration), which may result in an effective mobility edge 
and metal-insulator transition.

\section{\label{section:conclusions} Conclusions}

In summary, we have presented a systematical numerical study of Anderson localization 
in graphene with short-range disorder, using the real-space Kubo-Greenwood formalism and 
simulating uncorrelated Anderson disorder and vacancy defects. For graphene with Anderson disorder, 
the localization lengths for various quasi-one-dimensional systems with different 
widths $L_M$, disorder strengths, energies, edge types, and boundary conditions were calculated, 
and results for smaller systems were checked against the standard transfer matrix method with good agreement. 
We have found that the localization lengths $\lambda_M$ can be well described by a simple scaling function, 
$\lambda_M/L_M=\ln(1+k \xi/L_M)/k$, with $k$ being close or equal to $\pi$. Deviations from this scaling 
law occur due to finite-size effects, which manifest themselves when $L_M$ is comparable to or even smaller 
than the mean free path $l_{\textmd{e}}$. The two-dimensional localization lengths $\xi$ obtained using 
this scaling function are found to be consistent with the approximation based on
diffusive transport properties: $\xi=2l_{\textmd{e}}\exp[\pi\sigma_{\textmd{sc}}/G_0]$, 
where $\sigma_{\textmd{sc}}$ is the semiclassical conductivity and $G_0=2e^2/h$ is the conductance quantum. 
By calculating the 2D conductivity in the weak and strong localized regimes, with the finite-size
effects identified and eliminated by using sufficiently large simulation domain size, 
we also obtained a universal renormalization group $\beta$ function for 2D conductivity.
For graphene with vacancy disorder, we have demonstrated another finite-size effect in the 
real-space Kubo-Greenwood method, which occurs when the simulation cell length is 
not sufficiently large compared with $\lambda_M$. Surprisingly, the same scaling function
proposed based on the results for Anderson disorder also applies to graphene with vacancy
defects. The charge neutrality point in graphene with vacancy defects, however, 
exhibits an abnormally large peak value for the running conductivity in the ballistic-to-diffusive
regime. We have suggested that this abnormal behavior may be resulted form the local density 
of states caused by the resonant states located around the vacancy sites and presented
evidence that the charge neutrality point is exponentially localized. 
Our work thus suggests that the localization behavior of graphene with 
short-range disorder is to a large extent similar to conventional two-dimensional systems 
(such as the square lattice studied in the Appendix).

\begin{acknowledgments}
We thank A.-P. Jauho, K. L. Lee, D. Mayou, R. Mazzarello, S. Roche, 
R. A. R{\"o}mer, T.-M. Shih, and I. Zozoulenko for helpful discussions and comments. 
This research has been supported by the Academy of Finland through its Centres of Excellence 
Program (Project No. 251748). We acknowledge the computational resources provided by 
Aalto Science-IT project and Finland's IT Center for Science (CSC). 
\end{acknowledgments}

\appendix

\section{\label{section:square} Square lattice with Anderson disorder}

\begin{figure}
\begin{center}
  \includegraphics[width=\columnwidth]{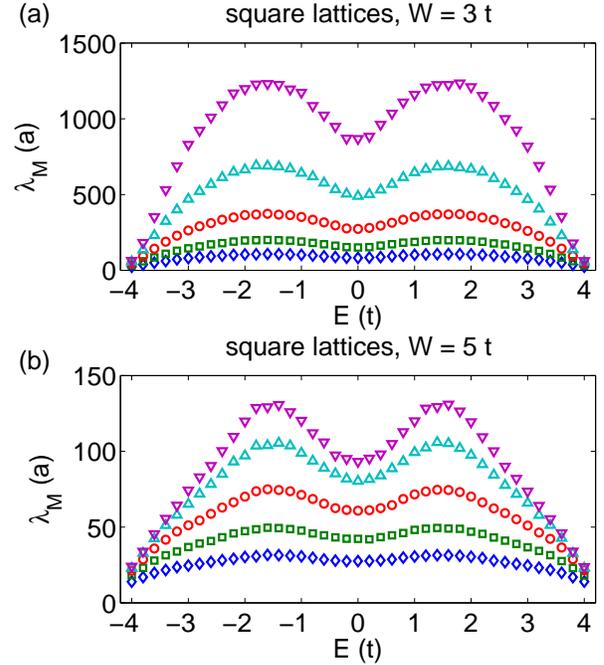}\\
  \caption{(color online) 
Q1D Localization length as a function of energy for square lattices with $W=3t$ (a) 
and $W=5t$ (b). The diamonds, squares, circles, upper triangles, and lower triangles 
correspond to $M=32$, 64, 128, 256, and 512, respectively. Free boundary conditions are applied 
along the transverse direction for the Q1D systems. Error bars are comparable to the
marker sizes and thus omitted.}
\label{figure:square_lambda}
\end{center}
\end{figure}

\begin{figure}
\begin{center}
  \includegraphics[width=\columnwidth]{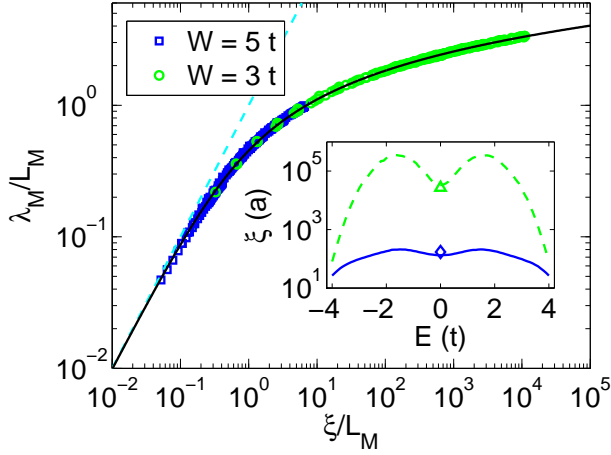}\\
  \caption{(color online) 
One-parameter scaling of localization length for square lattices 
with $W=3t$ and $W=5t$. The localization length divided by the width, $\lambda_M/L_M$, is plotted 
as a function of $\xi/L_M$, where $\xi$ is the 2D localization length obtained by fitting the data 
in Fig.~\ref{figure:square_lambda} against the scaling function.  The solid line represents 
the scaling function given by Eq.~(\ref{equation:scaling_function}) with $k=\pi$ and the dashed 
line represents the identity function $f(x)=x$. Note that $L_M=Ma$ for square lattice, where $a$ 
is the lattice constant. The inset shows the 2D localization length as a function of energy for 
$W=3t$ (dashed line) and $W=5t$ (solid line), with the triangle and diamond denoting the corresponding 
results for $E=0$ by Schreiber and Ottomeier \cite{schreiber1992}.}
\label{figure:square_xi}
\end{center}
\end{figure}

In this appendix, we show that the scaling function in Eq.~(\ref{equation:scaling_function}) 
with $k=\pi$ also applies to a square lattice with uncorrelated Anderson disorder, i.e., 
random on-site potentials uniformly distributed in an interval of $[-W/2, W/2]$. To this end, 
we first calculate the Q1D localization lengths using Eq.~(\ref{equation:xi_from_msd}). 
Figures~\ref{figure:square_lambda}(a) and \ref{figure:square_lambda}(b) show the results 
for $W=3t$ and $W=5t$, respectively. As can be seen from Fig.~\ref{figure:square_xi}, 
all the data with $32\leq M \leq 512$ are correlated by the scaling function very well, 
without any abnormal behavior resulting from the finite-size effect. Even the maximum 
mean free path for the square lattice with the weaker disorder strength, $W=3t$, is less than $10a$, 
which is well below the smallest value of $M$ considered. Therefore, all the data are in 
the scaling regime and follow the scaling curve. The obtained 2D localization lengths are shown 
in the inset, from which we see that the results for the band center are consistent with 
previous results by Schreiber and Ottomeier \cite{schreiber1992}. The results for other points 
away from the band center with $W=5t$ are also consistent with those by 
Zdetsis \textit{et al.} \cite{zdetsis1985}, exhibiting maximum values of $\xi$ around $E=\pm 2t$.

\begin{thebibliography}{99}
\bibitem{note_on__convention}
Note that there are different conventions for the definition of the localization length, 
which usually differ by a factor of 2. We have consistently followed the conventions widely 
used in the transfer matrix community. The reader should be aware of this when comparing 
our equations and results with others.
\end{thebibliography}
\end{document}